\RequirePackage{fix-cm}
\newcommand{\orcid}[1]{\href{https://orcid.org/#1}{\includegraphics[width=8pt]{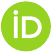}}}

\documentclass[twocolumn,epjc3]{svjour3}  
\smartqed  
\RequirePackage{graphicx}
\RequirePackage[numbers,sort&compress]{natbib}
\RequirePackage[colorlinks,citecolor=blue,urlcolor=blue,linkcolor=blue]{hyperref}
\RequirePackage{amsmath}
\RequirePackage{amssymb}
\RequirePackage{makecell}

\journalname{Eur. Phys. J. C}
\usepackage{siunitx}
\usepackage{xspace}

\newcommand{\micro}{\ensuremath{\mu}}

\newcommand{\bbonu}{\ensuremath{\beta\beta0\nu}}

\newcommand{\Qbb}{\ensuremath{Q_{\beta\beta}}}

\newcommand{\XE}{\ensuremath{{}^{136}\rm Xe}}

\newcommand{\TL}{\ensuremath{{}^{208}\rm{Tl}}}

\newcommand{\BI}{\ensuremath{{}^{214}\rm Bi}}

\newcommand{\batp}{\ensuremath{{\rm Ba^{2+}}}}
\newcommand{\nhfp}{\ensuremath{{\rm NH_4^{+}}}}
\newcommand{\nht}{\ensuremath{{\rm NH_3}}}

\newcommand{\Xe}[1]{\ensuremath{^{#1}\mathrm{Xe}}\xspace}

\DeclareSIUnit\c{\mbox{$c$}}
\DeclareSIUnit\magn{\mbox{$\times$}}
\DeclareSIUnit\min{min}
\DeclareSIUnit\week{week}
\DeclareSIUnit\year{yr}
\DeclareSIUnit\years{years}
\DeclareSIUnit\yr{yr}
\DeclareSIUnit\standard{std}
\DeclareSIUnit\str{sr}
\DeclareSIUnit\ppm{ppm}
\DeclareSIUnit\ppb{ppb}
\DeclareSIUnit\ppt{ppt}
\DeclareSIUnit\pe{PE}
\DeclareSIUnit\spe{SPE}
\DeclareSIUnit\ev{events}
\DeclareSIUnit\ct{counts}
\DeclareSIUnit\neutron{\mbox{$n$}}
\DeclareSIUnit\smp{samples}
\DeclareSIUnit\Sample{S}
\DeclareSIUnit\ch{ch}
\DeclareSIUnit\hit{hit}
\DeclareSIUnit\hits{hits}
\DeclareSIUnit\bin{(\mbox{5-PE}~bin)}
\DeclareSIUnit\sgm{\mbox{$\sigma$}}
\DeclareSIUnit\rms{RMS}
\DeclareSIUnit\keVr{\mbox{keV$_{\rm nr}$}}
\DeclareSIUnit\keVee{\mbox{keV$_{e{\rm e}}$}}
\DeclareSIUnit\ph{photon}
\DeclareSIUnit\pes{pes}
\DeclareSIUnit\el{electrons}
\DeclareSIUnit\pm{PMT}
\DeclareSIUnit\inch{"}
\DeclareSIUnit\bit{bit}
\DeclareSIUnit\sample{samples}
\DeclareSIUnit\barn{barn}
\DeclareSIUnit\bara{bar}
\DeclareSIUnit\barg{barg}
\DeclareSIUnit\mlardepth{\mbox(meter~of~\LAr~depth)}
\DeclareSIUnit\Curie{Ci}
\DeclareSIUnit\psi{psi}
\DeclareSIUnit\parsec{pc}
\DeclareSIUnit\liveday{\mbox{live-days}}
\DeclareSIUnit\days{\mbox{days}}
\DeclareSIUnit\day{\mbox{day}}
\DeclareSIUnit\miles{\mbox{miles}}
\DeclareSIUnit\degreeC{\mbox{$^{\circ}$C}}
\DeclareSIUnit\electron{\mbox{$e^-$}}
\DeclareSIUnit\Euro{\mbox{\euro}}
\DeclareSIUnit\cph{cph}
\DeclareSIUnit\neq{neq}
\DeclareSIUnit\unit{unit}
\DeclareSIUnit\byte{Byte}
\DeclareSIUnit\Bq{\becquerel}

\begin{document}

\title{A journey to ITACA
}
\subtitle{Ion Tracking with Ammonium Cations Apparatus}


\author{J.\,J.~G\'omez-Cadenas\orcid{0000-0002-8224-7714}\thanksref{e1,addra,addrb}
        \and
        L.~Arazi\orcid{0000-0002-7624-5827}\thanksref{addrc}
        \and
        M.~Elorza\orcid{0009-0000-4219-6193}\thanksref{addra,addrd}
        \and
        Z.~Freixa\orcid{0000-0002-2044-2725}\thanksref{addrd,addrb}
        \and
        F.~Monrabal\orcid{0000-0002-4047-5620}\thanksref{addra,addrb}
        \and
        A.~Pazos\orcid{0000-0002-6074-6213}\thanksref{addrd}
        \and
        J.~Renner\orcid{0000-0003-1843-2015}\thanksref{addre}
        \and
        S.R.~Soleti\orcid{0000-0002-5526-1414}\thanksref{e2,addra,addrb}
        \and
        S.~Torelli\orcid{0000-0003-3622-3524}\thanksref{addra}
}

\thankstext{e1}{e-mail: jjgomezcadenas@dipc.org}
\thankstext{e2}{e-mail: roberto.soleti@dipc.org}


\institute{Donostia International Physics Center, San Sebasti\'an / Donostia, E-20018, Spain \label{addra}
           \and
           Ikerbasque (Basque Foundation for Science), Bilbao, E-48009, Spain \label{addrb}
           \and
           Ben-Gurion University of the Negev, Beer-Sheva, 8410501, Israel \label{addrc}
           \and
           Universidad del País Vasco (UPV/EHU), San Sebasti\'an / Donostia, E-20018, Spain \label{addrd}
           \and
           Instituto de Física Corpuscular (IFIC), Paterna, E-46980, Spain \label{addre}
}

\date{Received: date / Accepted: date}

\maketitle

\begin{abstract}

A unique feature of gas xenon electroluminescent time projection chambers (GXeEL TPCs) in \bbonu\ searches is their ability to reconstruct event topology, in particular to distinguish “single-electron’’ from “double-electron’’ tracks, the latter being the signature of a \bbonu\ decay near the decay endpoint \Qbb. Together with excellent energy resolution and the t$_0$ provided by primary scintillation, this topological information is key to suppressing backgrounds. Preserving EL, however, requires operation in pure xenon (with helium as the only benign additive), where electron diffusion is large. Consequently, reconstructed track fidelity is limited by diffusion and intrinsic EL blurring.

We propose augmenting the detector with the ability to image not only the electron track but also the corresponding mirror ion track. Introducing trace amounts of \nht\ ($\sim$100~ppb) converts primary xenon ions into ammonium ions, \nhfp, via a fast two-step ion–molecule process involving charge transfer followed by proton transfer, while leaving EL unaffected. Electrons drift rapidly to the anode, producing the standard EL image, whereas \nhfp\ ions drift slowly toward the cathode, allowing time to determine the event energy and barycenter. For events in the region of interest, an ion sensor near the cathode at the projected barycenter captures the ions. Laser interrogation of the sensor’s molecular layer then reveals an ion-track image with sub-millimeter diffusion and no EL-induced smearing.

Combined electron–ion imaging strengthens topological discrimination, improving background rejection by about an order of magnitude and significantly extending the discovery potential of GXeEL TPCs for very long \bbonu\ lifetimes.

\end{abstract}

\section{Introduction}\label{s:intro}

\sloppy The observation of neutrinoless double-beta decay (\bbonu) would demonstrate that the neutrino is a Majorana particle, establish lepton-number violation, and shed light on the absolute neutrino mass scale~\cite{Avignone2008,Dolinski2019,Gomez-Cadenas:2023vca}. Detecting this rare process requires ultra-low background experiments with large isotopic masses and excellent energy resolution to resolve a possible signal peak at the decay Q-value (\Qbb) from background events.

The present generation of experiments has pushed the \bbonu\ sensitivity to half-lives beyond $10^{26}$~years, with the most stringent limit set by KamLAND-Zen at $T_{1/2} > 3.8 \times 10^{26}$ yr for $^{136}$Xe~\cite{KamLANDZen2024}. Other leading efforts include the LEGEND experiment, which has set a limit of $T_{1/2} > 1.9 \times 10^{26}$ yr for $^{76}$Ge~\cite{LEGEND2025}. The goal of next-generation experiments, being currently planned is to reach a sensitivity of  $T_{1/2} > 10^{27}$ yr, and, ultimately, $T_{1/2} > 10^{28}$ yr.

Over the last decade and a half, the technology of high pressure gas TPCs, operating with xenon and using electroluminescent amplification (GXeEL) has been developed by the NEXT collaboration, evolving i t from early prototypes~\cite{Alvarez:2011my,Alvarez:2012haa,Alvarez:2012hh,Alvarez:2012as,Alvarez:2012hu,Alvarez:2012xda} to a 100 kg-scale apparatus (NEXT-100)~\cite{NEXT:2025yqw}, currently searching for neutrinoless double beta decay processes (\bbonu) in \XE\ at the Canfranc Underground Laboratory. The large NEXT-100 detector achieves an excellent energy resolution of $(0.93 \pm 0.02)\%$ FWHM at \Qbb~\cite{nextcollaboration2025demonstrationsubpercentenergyresolution}, while the AXEL collaboration obtains $(0.67 \pm 0.08)$\% in a mid-sized prototype~\cite{Akiyama:2025dia}.


A GXeEL TPC provides a distinctive topological signature that enables identification of events with two emitted electrons. A \bbonu\ candidate in the apparatus is defined as a single, continuous track (with no detached energy deposits), reconstructed with total energy within the region of interest (ROI) around \Qbb\ and fully contained inside the fiducial volume, away from detector surfaces. This ``single-track'' requirement suppresses nearly all radioactive backgrounds, leaving only single-electron events with energies near \Qbb, such as those arising from photoelectric interactions of the 2.446 MeV $\gamma$ line of \BI\ and the 2.615 MeV line of \TL. The effectiveness of this topological discrimination has been demonstrated by the NEXT collaboration~\cite{Ferrario:2015kta}. 

However, the performance of the topological signature in a GXeEL TPC is limited by the use of pure xenon and by electroluminescent (EL) amplification. As ionization electrons drift through the dense gas, they undergo diffusion, producing a charge cloud whose transverse size increases with the drift length. 
In addition, EL amplification introduces further blurring, as the EL photons are emitted isotropically. 

The NEXT collaboration has extensively investigated methods to reduce electron diffusion in xenon gas. One of the most effective proposed approaches is the use of Xe/He mixtures (typically 85/15\%)~\cite{Felkai:2017oeq,NEXT:2019oxh,McDonald_2019}, which may reduce the transverse diffusion coefficient by about a factor 2. Other additives such as CH$_4$ and CO$_2$ have also been studied~\cite{Henriques:2018tam,Henriques:2017rlj}. These gases cool the drifting electrons efficiently but strongly quench xenon scintillation, creating a trade-off between reduced diffusion and loss of electroluminescent (EL) yield. 

Here we propose a new concept to measure event topology in an GXe TPC, which we assumed EL, given that $t_0$ is needed and excellent energy resolution is a bonus. The method relies on recording two complementary tracks for each event: the {\em electron track (eT)}, the established signature in a GXe TPC, and an {\em ion track (iT)}, formed by suitable positive ions that can be trapped by a molecular layer of sensors similar to those developed by the NEXT collaboration in the context of their ``barium tagging'' program ~\cite{Jones:2016qiq,McDonald:2017izm,Thapa:2019zjk,rivilla_fluorescent_2020,doi:10.1021/acssensors.4c01892,ZF8546,ZF7462,ZF7997,NEXT:2022ita}. It is worth noting that related concepts based on negative ion drift, originally developed by the DRIFT collaboration for directional dark matter searches to strongly suppress diffusion, have been explored previously and demonstrated experimentally, albeit in a different physics context~\cite{Martoff:2000wi,Snowden-Ifft:1999reu}. More recently, similar ideas have been revisited and extended in the context of combined ion–electron imaging~\cite{jones2022ionfluorescencechamberifc}.
We call this concept ITACA (Ion Tracking with Ammonium Cations Apparatus).


This paper is organized as follows: section \ref{sec:toplogy} reviews the reconstruction of the event topology in a GXeEL TPC. Section \ref{sec:itaca} proposes a conceptual design for the ITACA detector. The principle of operation is examined in \ref{sec:po}. In section \ref{sec:itd} we describe the ion track detector and in section \ref{sec:sensi} we briefly discuss the potential enhancement in sensitivity provided by the ITACA technique. 
%
We present our conclusions in 
 \ref{sec:conclu}.

\section{Reconstruction of event topology in a GXeEL TPC}
\label{sec:toplogy}

\begin{figure}[!htbp]
    \centering
    \includegraphics[width=1\linewidth]{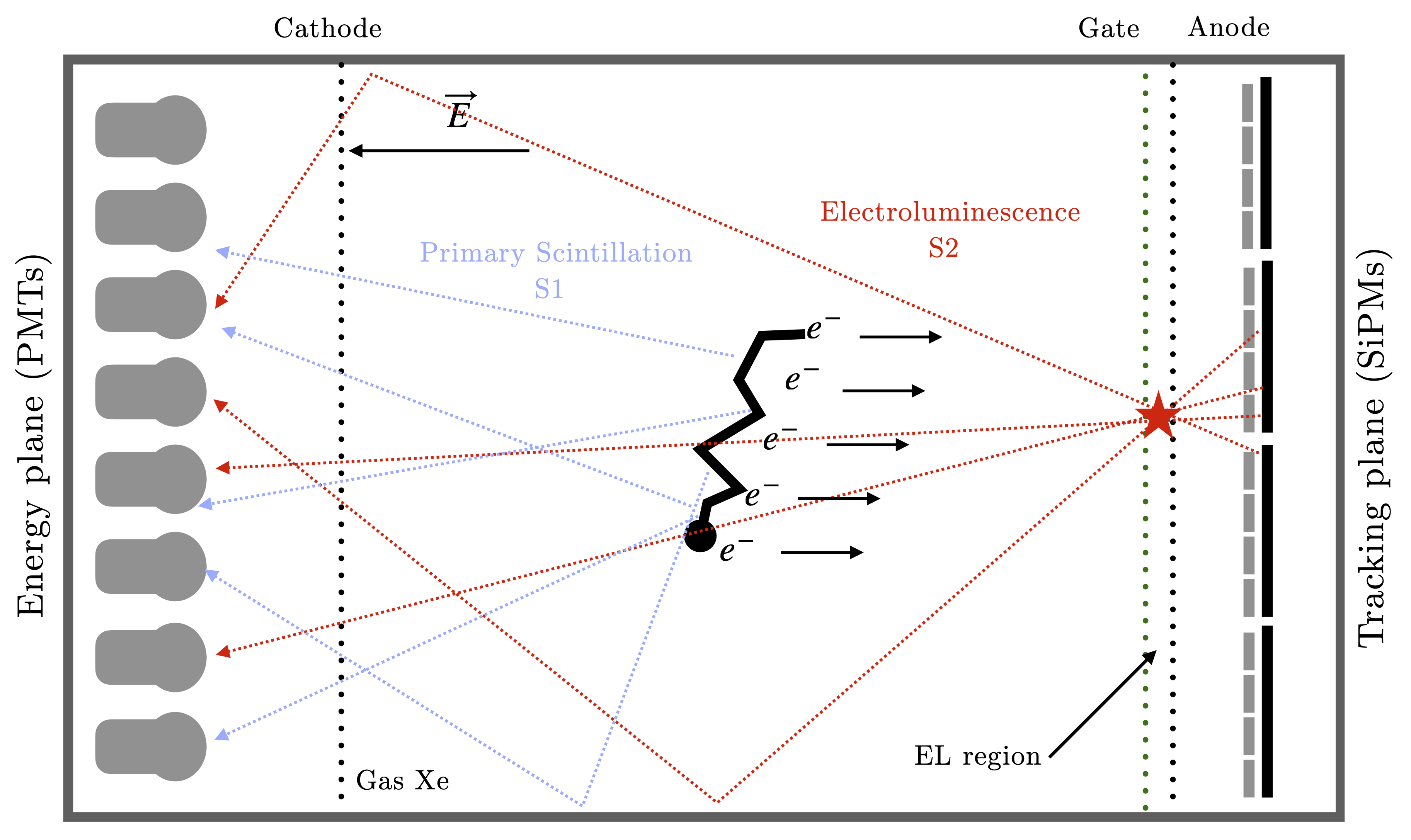}
    \caption{Principle of operation of of an asymmetric GXeEL TPC with SiPMs and PMTs.}
    \label{fig:principle}
\end{figure}

A GXeEL TPC operates as shown in Figure \ref{fig:principle}. Xenon emits UV scintillation at 172 nm due to excimer formation and decay, marking the event start, $t_0$. A moderate electric field (a few hundred V/cm) transports ionization electrons to the TPC anode, crossing a higher electric field (EL) region. Here, electrons excite xenon atoms without further ionization, causing a proportional burst of secondary scintillation light. This amplification, with sub-poissonian fluctuations ensures excellent energy resolution by consistently collecting almost all ionization charge.

The EL region is defined by two transparent grids separated by a small gap (about 5-10 mm): the anode, held at ground potential, and the gate, biased to produce a reduced electric field of roughly 2 kV/cm/bar. Under typical conditions, about $10^3$ photons are emitted per ionization electron. Photons directed toward the anode are detected in a pixelated detector. All the NEXT detectors, as well as the AXEL prototypes use an array of silicon photomultipliers (SiPMs) with a typical pitch of 10 mm. Event tracks are reconstructed by combining the spatial information from the SiPMs with the timing provided by the TPC. In an asymmetric TPC, such as the NEXT-100 detector and the AXEL prototype, photons emitted toward the cathode are recorded by photomultipliers located behind the transparent cathode, which also detect the primary scintillation signal (S1). 
The decay
$
\mathrm{^{136}Xe \;\to\; ^{136}Ba^{2+} + 2e^-}
$
releases $Q_{\beta\beta}=2.458~\mathrm{MeV}$, and thus 
$1.1\times10^5$ electron-ion pairs, distributed along a twisted trajectory of about 10-15 cm path length when operating at pressures around 15 bar.


\begin{figure*}[!htbp]
    \centering
    \includegraphics[scale=0.40]{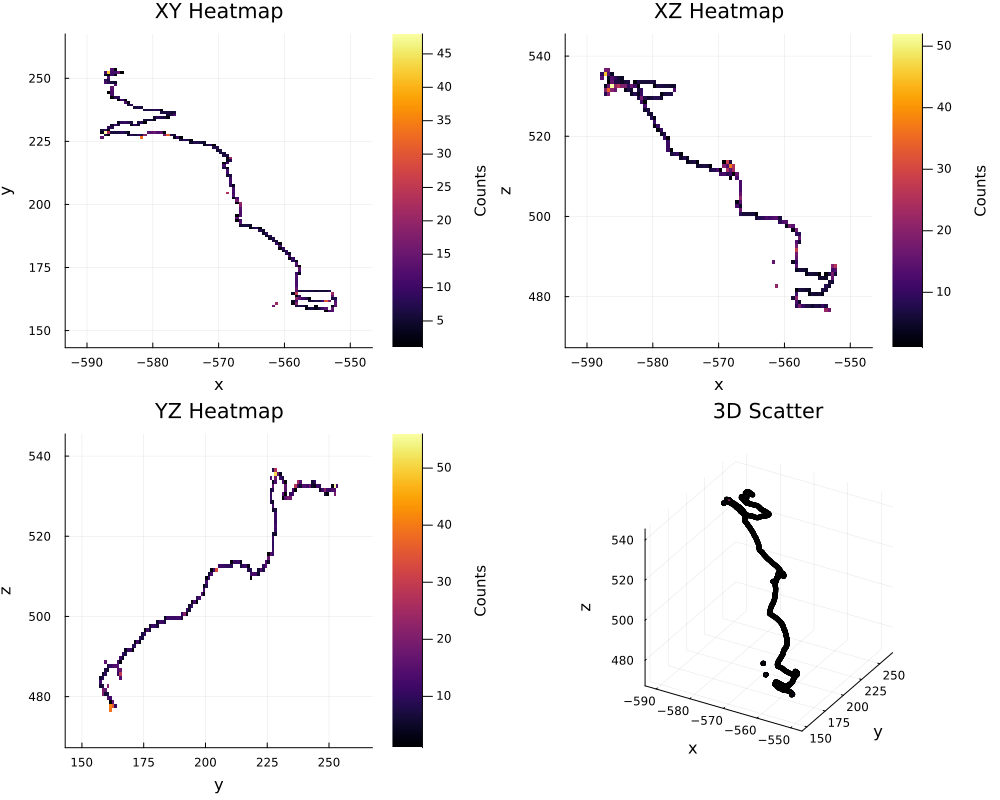}
    \caption{A simulated \bbonu\ track in HPXe.}
    \label{fig:mctrack}
\end{figure*}

Figure~\ref{fig:mctrack} displays the ionization pattern from two electrons generated in a simulated \bbonu\ event. Multiple scattering bends their paths, which end in dense ``blobs" of energy deposition at the Bragg peaks. 

\begin{table}[h!]
\centering
\caption{Measured longitudinal ($D_L$) and transverse ($D_T$) diffusion coefficients 
for pure xenon and 10\% Xe/He gas mixtures, expressed in $\sqrt{\mathrm{bar}}\,\mathrm{mm}/\sqrt{\mathrm{cm}}$.}
\begin{tabular}{lcc}
\hline
\textbf{Gas mixture} & \makecell{ $D_L$\\($\sqrt{\mathrm{bar}}\,\mathrm{mm}/\sqrt{\mathrm{cm}}$) }  & \makecell{ $D_T$ \\($\sqrt{\mathrm{bar}}\,\mathrm{mm}/\sqrt{\mathrm{cm}}$)} \\
\hline
Pure Xe      & 0.900 & 3.500 \\
Xe + 10\% He & 0.750 & 1.600 \\
\hline
\end{tabular}
\label{tab:diffusion}
\end{table}

The reconstruction of topology in a GXeEL TPC is influenced by two distinct smearing effects. The first arises from the spatial spread of electroluminescent (EL) light as electrons are accelerated in the EL region. This effect can be accurately modeled by a Gaussian distribution and precisely characterized using calibration data~\cite{Simon:2017pck,Simon:2018vep}. The second, and more significant, contribution comes from the longitudinal and transverse diffusion of drifting electrons, which can be expressed as: 
\begin{equation}
\sigma_{l,t} ({\rm mm}) = \frac{D_{l,t}}{\sqrt{P} } \times \sqrt{L}
\end{equation}
where, if $L$ is expressed in cm and $P$ in bar, $D_{l,t}$ is expressed in 
$\mathrm{\sqrt{bar}\ mm /\sqrt{cm}}$. 

Table \ref{tab:diffusion} shows the values of ${D_{l,t}}$ for pure Xenon and for a 0.9/0.1 mixture of Xenon and Helium, which substantially reduces the transverse diffusion~\cite{Felkai:2017oeq,Henriques:2018tam,McDonald_2019}.

\begin{figure*}[!htbp]
    \centering
    \includegraphics[width=0.99\linewidth]{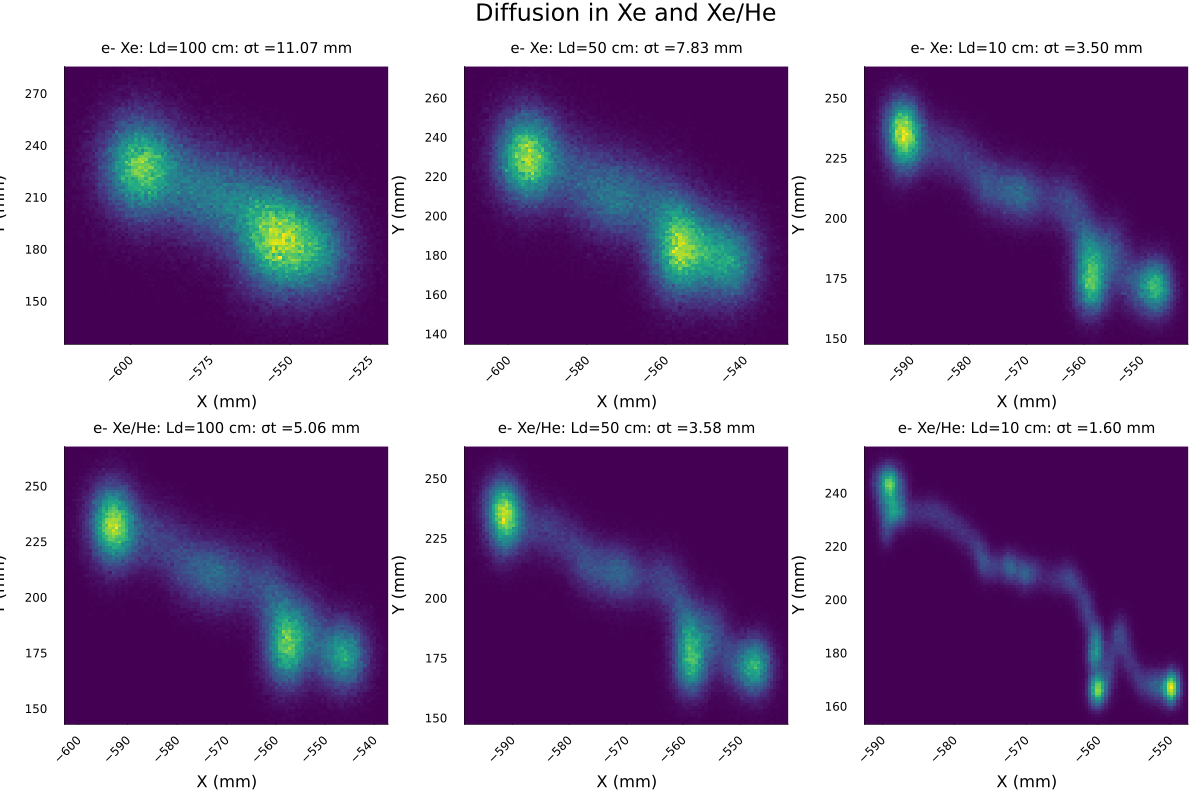}
    \caption{Reconstruction of a \bbonu\ track a GXeEL TPC with 1 m electron drift. The upper row shows the XY projection of  the track in pure xenon, for the characteristic distances. Near the cathode ($\sim 100$ cm)
    in the middle of the drift ($\sim 50$ cm) and near the anode ($\sim 10$ cm). The bottom row shows the same projections for Xe/He. }
    \label{fig:etrack}
\end{figure*}

Figure~\ref{fig:etrack} shows the reconstruction of the XY projection of the event displayed 
in Figure ~\ref{fig:mctrack} for the case of pure xenon and for a 0.9/0.1 Xe/He mixture. Three representative drifting distances are considered. Notice that the smearing of the tracks near the cathode is very large, even with Xe/He mixture. As the impact of the diffusion is reduced, the details of the track becomes sharper, ultimately revealing the two blobs (as well as any potential energy deposit near the track, such as for example the characteristic 30 keV X-ray in the photoelectric interaction of the 2448 MeV $\gamma$ produced in $^{214}$Bi decays) with great detail. 

To separate signal and background events in the search for \bbonu\ decays, a GXe TPC exploits the different topology of the two electrons emitted in a double beta decay, and the single electrons (often accompanied by additional energy depositions) which constitute the main backgrounds. The three main sources of background are:  the $\gamma$ line of 2447 keV arising from \BI\ decays, ($\gamma_{\BI}$); the $\gamma$ line of 2615 keV arising from \TL\ decays ($\gamma_{\TL}$), and the single electrons from the \Xe{137} $\beta$ decay with energies compatible with the region-of-interest (ROI) of the \bbonu\ search (typically 1 FWHM around \Qbb).  Given that the ROI width for EL detectors if of the order of 15-25 keV, while the distance between \Qbb\ and the $\gamma_{\BI}$ photopeak is 10 keV, topology becomes specially important to suppress this background source, but is also needed to reduce the flat background from \Xe{137} decays and Compton interactions of the $\gamma_{\TL}$.

The $\gamma_{\BI}$ photopeak produces a single electron, accompanied 85\% of the times by the characteristic de-excitation X-ray of xenon (with an energy of about 30 keV), and occasionally Bremsstrahlung emission.  The  $\gamma_{\TL}$~introduces background in the ROI via Compton electrons, which come together with the scattered Compton photon and eventually Bremsstrahlung. The \Xe{137} $\beta$ decay produces a single electron, with end point energy of 4.17 MeV, thus intersecting the ROI. Also this electron can emit Bremsstrahlung. In all cases, therefore, imposing a single track with no extra energy deposition is advantageous to separate signal from backgrounds. The second requirement is to be able to separate the signature of one electron from that of two electrons. Thus, the signature of the double beta decay is a single connected track with two energy blobs of similar intensity in each of the track extremes, as those that can be observed in Figure \ref{fig:etrack}. 

Reducing diffusion has two positive effects. On one side, it permits a better separation between  deposits (such as the 30 keV X-ray signaling a $\gamma_{\BI}$ photoelectric interaction) and the main electron track. On the other, it allows resolving better the two blobs of a signal event.

\section{The ITACA Detector}
\label{sec:itaca}

\begin{figure}[!htbp]
    \centering
    \includegraphics[width=1\linewidth]{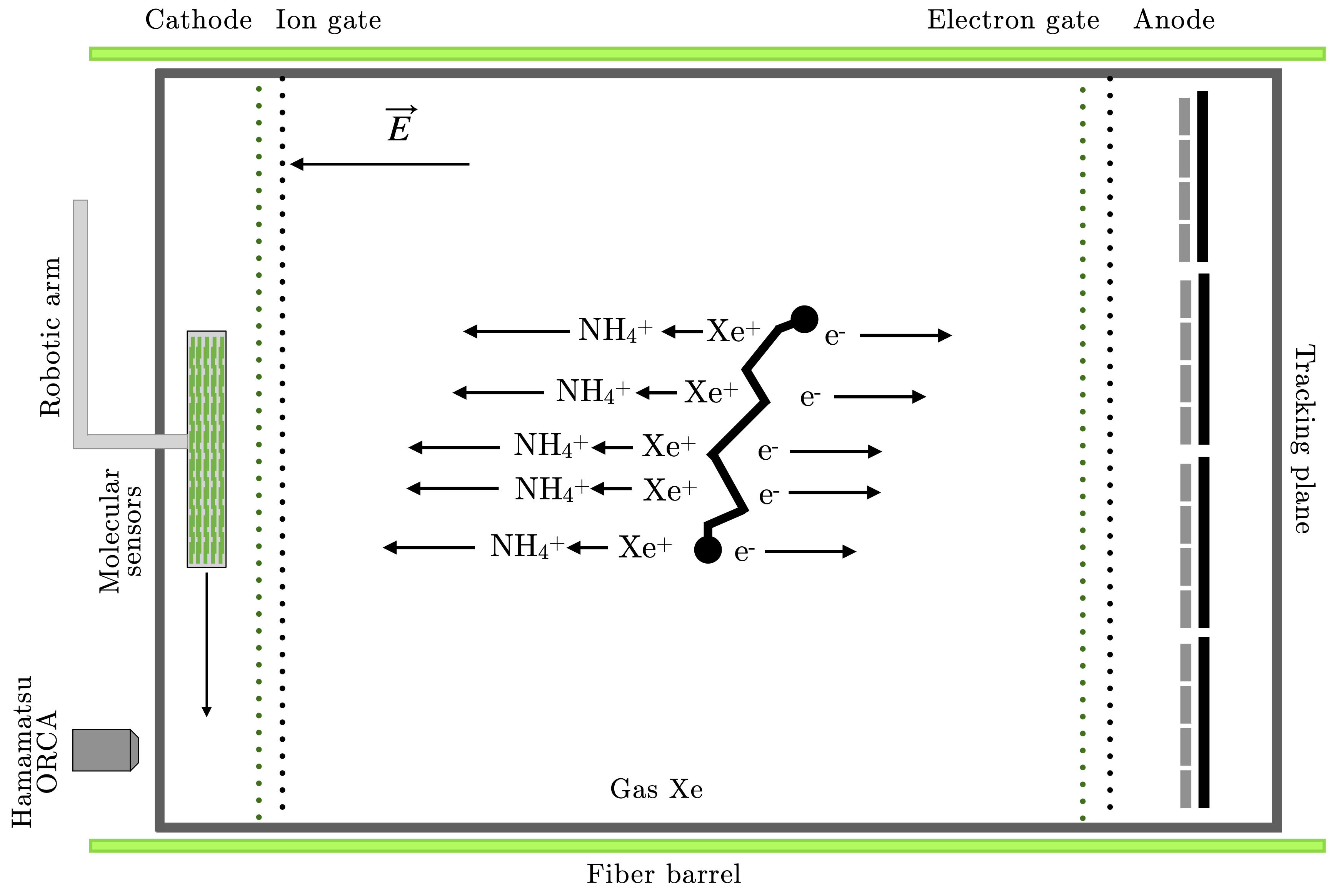}
    \caption{Principle of operation of ITACA. }
    \label{fig:itaca}
\end{figure}

 The ITACA (Ion Tracking with Ammonium Cations Apparatus) concept is shown in Figure \ref{fig:itaca}. Trace amounts (100 pb) of ammonia (\nht) are added to the gas (assumed to be a 0.9/0.1 Xe/He mixture\footnote{So far, all GXeEL TPCs have operated with pure xenon, although the NEXT collaboration aims to use eventually Xe/He mixtures~\cite{NEXT:2020amj}.}. In spite of the low concentration, this is enough to transform positive xenon ions into ammonium (\nhfp), without quenching the EL light or affecting the electron drift. 
 
  \begin{figure*}[htbp!]
      \centering
      \includegraphics[width=0.99\textwidth]{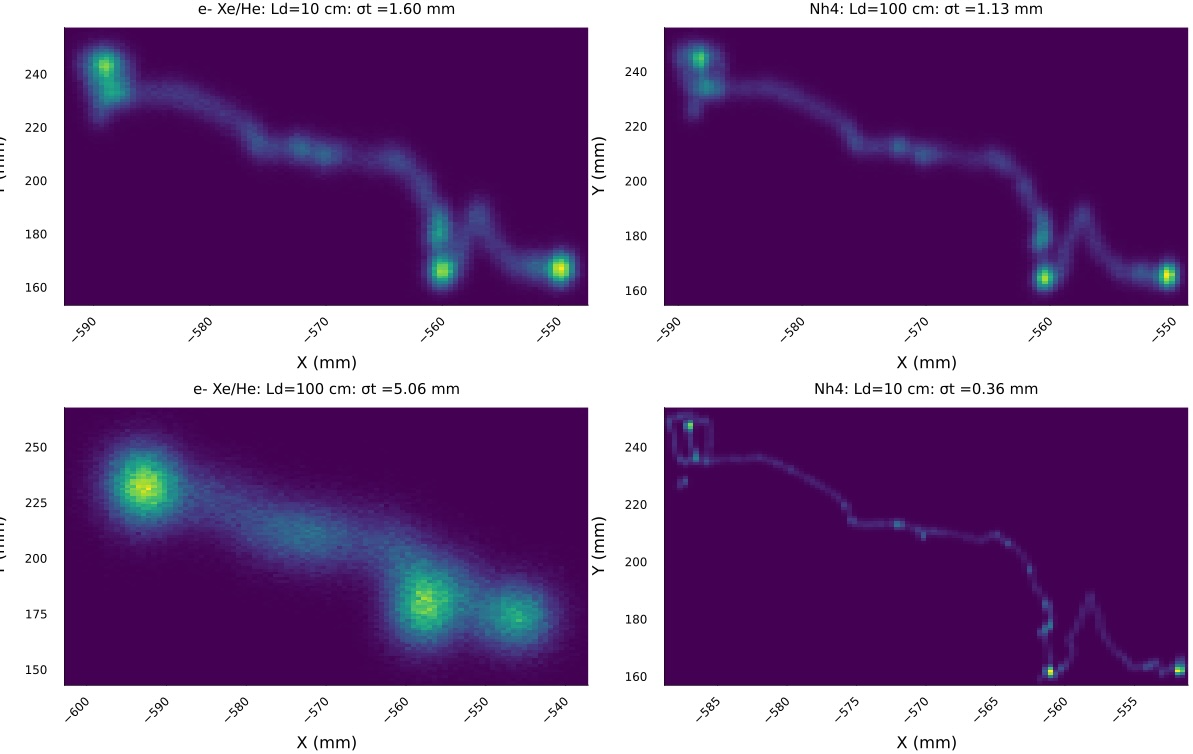}
      \caption{The diffusion in ion and electron tracks is anticorrelated. The upper panel shows the case in which the electrons are near the anode (10 cm), and thus ions are far away from the cathode (100 cm); the bottom panel shows the inverse case. Ions are close to the cathode (10~cm) and thus electrons are far from the anode (100 cm). The availability of the ion track allows a uniform topological reconstruction across the detector. }
      \label{fig:ctracks}
  \end{figure*}
  
Positive \nhfp\ ions, then, drift to the cathode, in a time scale of seconds, while the electrons drift to the anode three orders of magnitude faster. Ion diffusion is much smaller than electron diffusion, and thus, reconstruction of the ion track (iT), combined with the electron track (eT) improves the topological rejection of backgrounds. This is enhanced by the fact that the
diffusion in the ion and electron tracks is anticorrelated, as illustrated in Figure  \ref{fig:ctracks}. Consider two extreme cases. When electrons are near the anode (10 cm) diffusion is small, and the reconstructed track can be combined with the ion track. Instead, when electrons are far from the anode (100 cm) the reconstructed electron track is totally blurred, but the ion track is reconstructed with very small diffusion, and the resulting track resolves the blobs in exquisite detail. The availability of the ion track, thus, allows a uniform topological reconstruction across the detector.

For definiteness, let us consider an implementation of the ITACA concept able to hold 1 tonne of xenon at high pressure, with the following parameters:
\begin{enumerate}
\item TPC diameter: 200 cm.
\item TPC length : 200 cm.
\item Operating pressure: 30 bar.
\item Mass of xenon: 1.2 ton.
\item Density: $n_\mathrm{{Xe}} = 9 \times 10^{20}~ \mathrm{atoms/cm^3}$.
\item Drift field: 400 V/cm.
\end{enumerate}

ITACA is an asymmetric EL TPC, like NEXT-100, but instrumented like the proposed NEXT-HD detector~\cite{NEXT:2020amj}: it deploys a fiber barrel calorimeter~\cite{Soleti:2023sac}, able to measure the primary scintillation S1 (and thus the start-of-the-event t$_0$) as well as the EL light S2, and a Dense Silicon Plane, able to measure S2 and to reconstruct the electron track. 

 Operation at higher pressure results in shorter electron tracks (w.r.t. NEXT-100 and NEXT-HD), but also smaller diffusion (a factor $\sqrt{2}$ less than NEXT-100 and NEXT-HD). Electron drift velocity is of the order of 100 cm/ms, while ion drift velocity is of the order of 25 cm/s, as we will discuss below. Operation of ITACA at this high pressure has the advantage of allowing a more compact detector and permits a longer fiducial region. 
 
 This last point deserves some discussion. To define the fiducial region in $z$, we need to assess what is the minimum distance between the barycenter of a \bbonu\ decay candidate and the anode that allows the reconstruction of the ion track. We allow a minimum delay time of $t_d \sim t_0 + 500\mathrm{~ms}$, which is needed to: a) compute the energy of the event; b) reconstruct the event rough topology and compute its barycenter, $P$; c) compute the projected position of the barycenter, $P'$ in the anode detection plane; d) move the so-called Molecular Ion Detector, defined in section \ref{sec:itd}, to $P'$, so that it can capture the incoming ion track. 
 
 On the other hand, the ion drift velocity is 20 cm/s, thus allowing 10 cm minimum drift is sufficient to grant $\mathrm{t_d > 500~ms}$. The fiducial region is thus 190 cm, and only 5\% of the fiducial volume in $z$ is lost. 

 Ion collection is performed through a gated scheme, in which the ion detector is exposed only within a selected time window following a validated event. This ion gate allows the collection process to be synchronized with the prompt electron signal and provides the basis for suppressing pile-up from unrelated ionization events.
 
\section{Principle of operation}
\label{sec:po}
\subsection{Ion Chemistry and Charge Transfer}\label{sec:ion_chemistry}

Primary ionization in xenon produces Xe$^+$, which rapidly converts to Xe$_2^+$ through
three-body association:
\begin{equation}
\mathrm{Xe^+ + 2Xe \rightarrow Xe_2^+ + Xe}.
\end{equation}

Trace amounts of ammonia convert the primary xenon ions into ammonium ($\mathrm{NH_4^+}$) through a rapid two-step mechanism: first, charge transfer from $\mathrm{Xe_2^+}$ to $\mathrm{NH_3}$ forming $\mathrm{NH_3^+}$, followed by proton transfer with a second $\mathrm{NH_3}$ molecule to produce \nhfp\:

\begin{align}
\label{eq:charge_transfer}\mathrm{Xe_2^+ + NH_3} &\rightarrow 2\,\mathrm{Xe} + \mathrm{NH_3^+},\\
\label{eq:proton_transfer}\mathrm{NH_3^+ + NH_3} &\rightarrow \mathrm{NH_4^+ + NH_2}.
\end{align}

\sloppy Laboratory data show that both processes are nearly collision-limited,
with rate constants $k_{\rm ct} \approx 6 \times10^{-10}~\mathrm{cm^3\, molecule^{-1}\, s^{-1}}$
for the Xe$_2^+$ + NH$_3$ charge-transfer channel~\cite{Levandier2010} and $k_{\rm pt} \approx 1.8 \times 10^{-9}~\mathrm{cm^3\, molecule^{-1}\, s^{-1}}$
for the proton transfer step~\cite{Conaway1987}. 
%
Both reactions are exothermic by roughly 1–2~eV.

Let us consider an addition of a mole fraction $f=100$~ppb of \nht\ to the xenon. 
At $T\simeq 20~^{\circ}$C and $P=30$~bar the xenon number density is 
$n_{\mathrm{Xe}} \approx 9 \times 10^{20}\ \mathrm{atoms/cm^{3}}$, 
therefore 
$n_{\nht} =f \cdot n_{\mathrm{Xe}} \approx 9\times10^{13}\,  \mathrm{atoms/cm^{3}}$.
Thus, for the charge–transfer step of eq.~\eqref{eq:charge_transfer} we obtain:
\[
\tau_{ct} \;=\; \frac{1}{k_{ct} \cdot n_{\nht}}
\;\approx\; \frac{1}{(6\times10^{-10})(9\times10^{13})}
\;\simeq\; 18.5~\si{\micro\second}\,,
\]
while for the proton–transfer step of eq.~\eqref{eq:proton_transfer} we obtain a reaction time:
\[
\tau_{\rm pt} \;=\; \frac{1}{k_{\rm pt} \cdot n_{\rm NH_3}}
\;\approx\; \frac{1}{(1.8\times 10^{-9})(9\times10^{13})}
\;\simeq\; 6~\si{\micro\second}\,,
\]

Thus, both steps are completed in less than 100~\si{\micro\second} at $5\tau$, which is
orders of magnitude shorter than ion drift times, ensuring near-complete conversion of the positive charge to \nhfp\
before arrival at the cathode. Notice that full conversion will happen also very fast with much smaller amounts of \nht\ (e.g., 10 ppb will convert the xenon ions in about 100 $\mu$s, while conversion with 1 ppb only requires 1 ms, still three orders of magnitude smaller than drift time). It follows that one has considerable tolerance in the amount of ammonia added to the detector. 

\subsection{Transport Properties}

The ion mobility $\mu$ is defined as the proportionality constant between the 
drift velocity $v_d$ and the electric field $E$:
\begin{equation}
v_d = \mu E.
\end{equation}
In the low-field (thermal) regime, $\mu$ 
depends on the ion–neutral collision frequency 
can be expressed~\cite{Mason1988} as
\begin{equation}
\mu = \frac{3q_e}{16N}
\sqrt{\frac{2\pi}{m_r k_B T}}
\frac{1}{\Omega(T)},
\label{eq:mason}
\end{equation}
where $q_e=1.602\times10^{-19}$~C, $N$ is the gas number density, and $m_r$ the ion–neutral reduced mass,
and $\Omega(T)$ the momentum-transfer collision integral.
The neutral polarizability $\alpha$ determines the strength of the long-range polarization potential
$V(r)=-\tfrac{1}{2}\alpha e^2 r^{-4}$.

\subsubsection*{Mobility in Argon at STP}

From~\cite{ABEDI2014101}, the reduced mobility for NH$_4^+$ in Ar is
$K_0 \approx 2.28 \pm 0.1~\mathrm{cm^2\,V^{-1}\,s^{-1}}$ at STP.
The corresponding mobility at 20~$^\circ$C and 1~bar is:
\begin{align}
\mu(20~^{\circ}\mathrm{C},1~\mathrm{bar}) 
= K_0 \frac{T}{T_0} \frac{P_0}{P}
&= (2.28 \pm 0.1)\times\frac{293}{273}\\
&\approx 2.5 \pm 0.11~\mathrm{cm^2\,V^{-1}\,s^{-1}}.\nonumber
\end{align}

\subsubsection*{Extrapolating from Argon to Xenon at the same $(T,P)$}

For interactions dominated by polarization forces,
the momentum-transfer integral scales as $\Omega(T)\propto(\alpha/T)^{1/2}$,
yielding the approximate scaling:
\begin{equation}
\frac{\mu_B}{\mu_A}\simeq
\sqrt{\frac{m_{r,A}}{m_{r,B}}}
\sqrt{\frac{\alpha_A}{\alpha_B}},
\end{equation}
where $m_r$ is the ion–neutral reduced mass and $\alpha$ the neutral polarizability.

Using standard gas-phase polarizabilities~\cite{mason1988tables}:
\begin{align}
\alpha_{\mathrm{Ar}} &= 1.6411~\text{\AA}^3, &
\alpha_{\mathrm{Xe}} &= 4.044~\text{\AA}^3,
\end{align}
gives $\alpha_{\mathrm{Ar}}/\alpha_{\mathrm{Xe}}\approx0.41$.
For NH$_4^+$, $m_i=18$~u, while $m_g^{\mathrm{Ar}}=40$~u and
$m_g^{\mathrm{Xe}}=131$~u, giving
$m_{r,\mathrm{Ar}}\approx12.4$~u and $m_{r,\mathrm{Xe}}\approx15.8$~u,
or $m_{r,\mathrm{Ar}}/m_{r,\mathrm{Xe}}\approx0.78$.
The mobility ratio is then $\mu_{\mathrm{Xe}}/\mu_{\mathrm{Ar}}\approx0.57$.

At 20~$^{\circ}$C~and 1~bar:
\begin{equation}
\mu^{\mathrm{Xe}}_{1\,\mathrm{bar}}
\approx 0.57\times(2.50\pm0.11)
\approx 1.42\pm0.06~\mathrm{cm^2\,V^{-1}\,s^{-1}}.
\end{equation}

\subsubsection*{Pressure scaling}

In the low-$E/N$ regime, $\mu \propto 1/N \propto 1/P$ (for fixed $T$), hence:
\begin{equation}
\mu^{\mathrm{Xe}}_{30\,\mathrm{bar}}
\approx \frac{\mu^{\mathrm{Xe}}_{1\,\mathrm{bar}}}{30}
\approx (0.047\pm0.001) ~\mathrm{cm^2\,V^{-1}\,s^{-1}}.
\end{equation}

\subsubsection*{Drift velocity}

For $E=400~\mathrm{V\,cm^{-1}}$:
\begin{equation}
v_d = \mu E \approx 0.05 \times400
\approx 20~\mathrm{cm\,s^{-1}},
\end{equation}
so drifting from the edge of the fiducial region (20 cm from the cathode) requires $\approx 0.5$~s, while drifting the full length of the chamber would take 5 seconds. 

\subsubsection*{Diffusion}

  \begin{figure}[htbp!]
      \centering
      \includegraphics[width=0.49\textwidth]{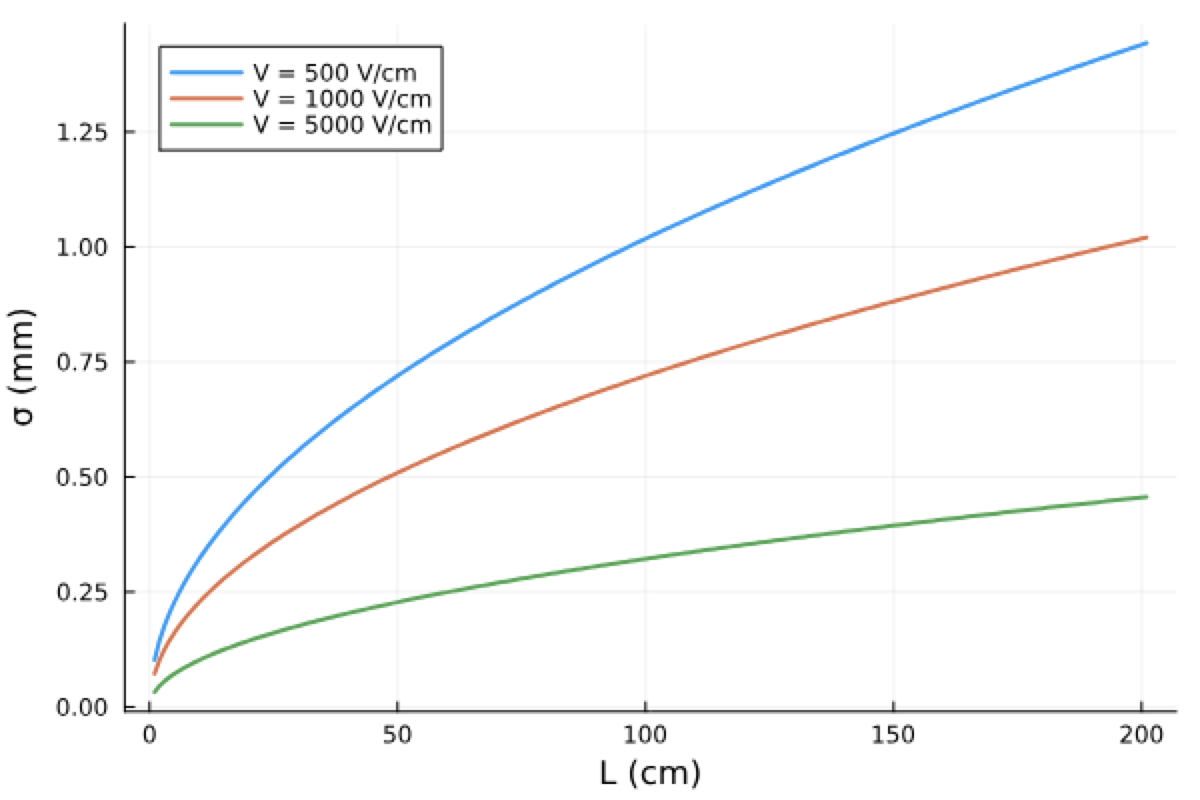}
      \includegraphics[width=0.49\textwidth]{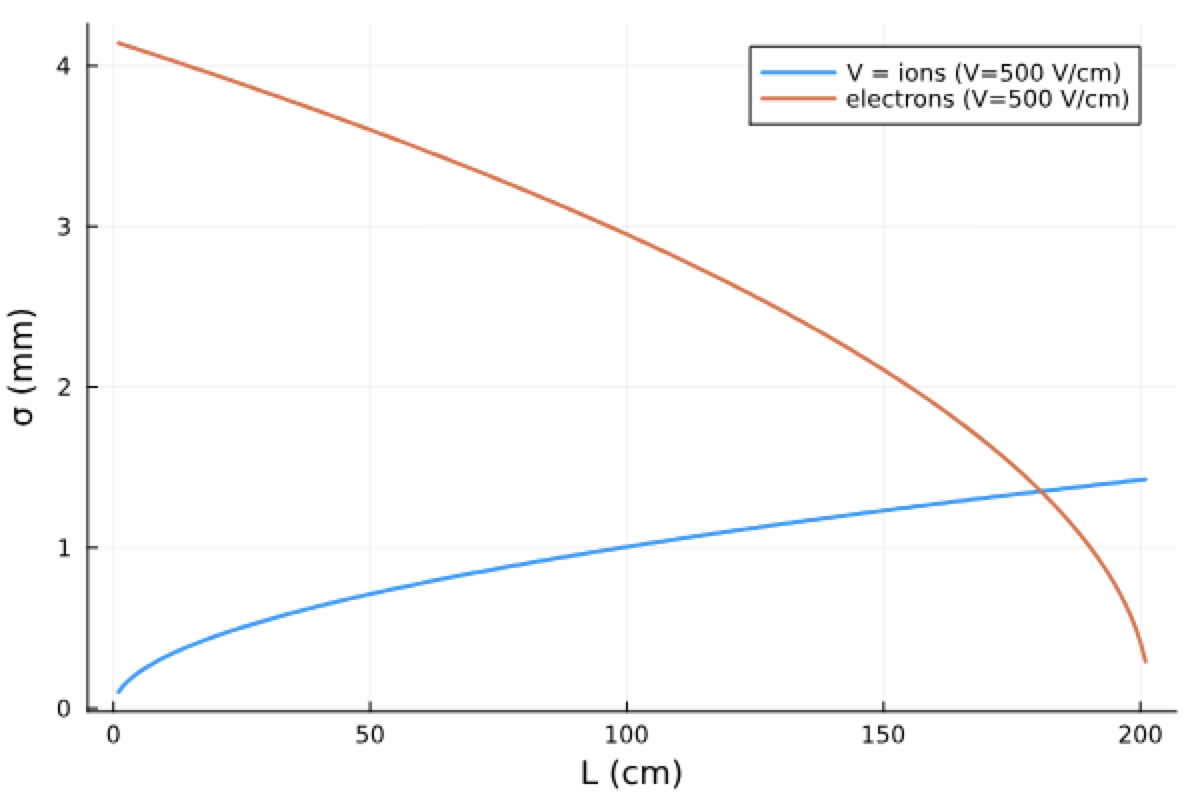}
      \caption{Top: diffusion of the ion track as a function of L for three different drift fields. Bottom: the diffusion of ions and electrons is anti-correlated (L is the distance to the node).}
      \label{fig:sigmat}
  \end{figure}

Assuming ions remain thermal, $D=\mu k_BT/q_e$,
\[
\sigma=\sqrt{2Dt}=\sqrt{\tfrac{2k_BT}{q_e}\tfrac{L}{E}} 
\]

Figure \ref{fig:sigmat} (left) shows the diffusion of the ion track as a function of the drift length for three different drift fields. Notice that a drift field of 500~V/cm requires a potential difference between the anode and the gate of 100~kV, a high, but achievable value at large pressure, where the buffering of the noble gas is very large. A drift field of 1000~V/cm would be very challenging, while larger drift fields would require substantial technological developments. 

Figure \ref{fig:sigmat} (right) shows the anti-correlation between the electron and ion tracks, where $L$ is the distance to the anode. Notice that for most of the drift length, the ion track has a much smaller diffusion than the electron track.  However, near the anode, both tracks can be combined to further reduce the diffusion. 

\subsection{VUV Absorption in Primary and Secondary Scintillation}

The Xe$_2^*$ excimer responsible for both primary (S1) and electroluminescent (S2)
emission decays predominantly at 172~nm.
Ammonia (NH$_3$) absorbs strongly in this region, with cross sections
$\sigma_{\mathrm{abs}}\sim 10^{-18}~\mathrm{cm^2}$~\cite{Chen1999_NH3_UV_PSS}
corresponding to dissociative excitation.
Both S1 and S2 photons can, in principle, be absorbed by trace NH$_3$ admixtures.

Primary scintillation photons are emitted throughout the TPC volume and may 
traverse meter-scale distances before reaching the photosensors.
Electroluminescent photons emitted backward follow similar paths.

The attenuation length for VUV photons in Xe with a trace NH$_3$ admixture is
\begin{equation}
\lambda_{\mathrm{abs}}=\frac{1}{n_{\mathrm{NH_3}}\sigma_{\mathrm{abs}}}
=\frac{1}{f\,n_{\mathrm{Xe}}\,\sigma_{\mathrm{abs}}}.
\end{equation}

Plugging in  $f=100$~ppb, $n_{\mathrm{Xe}}\approx 9\times10^{20}~\mathrm{atoms/cm^{-3}}$, then $n_{\mathrm{NH_3}}\approx 9\times10^{13}~\mathrm{molecules/cm^{-3}}$.
For a representative NH$_3$ cross section at 172~nm
$\sigma_{\mathrm{abs}} \approx 10^{-18}~\mathrm{cm^2}$:
\begin{equation}
\lambda_{\mathrm{abs}} \;=\;
\frac{1}{(9\times10^{13})(10^{-18})}
\;\approx\; 1.1\times10^{4}\ \mathrm{cm}.
\end{equation}

For a 200~cm path,
\begin{equation}
\frac{I(L)}{I_0}=\exp\!\left(-\frac{L}{\lambda_{\mathrm{abs}}}\right)
=\exp\!\left(-\frac{200}{1.1\times10^{4}}\right) \sim 0.98,
\end{equation}
resulting in a tiny loss of light at the maximum attenuation. For smaller concentrations (e.g, 10 ppb) the loss of light is totally negligible. 

\subsection{Electron Attachment in Xe--NH$_3$ Mixtures}

A possible concern when introducing molecular additives into xenon gas
is electron attachment, which could shorten the electron lifetime and degrade
the charge collection efficiency.

Electron attachment to NH$_3$ proceeds mainly via 
\emph{dissociative electron attachment} (DEA) resonances at
electron energies of approximately 5.5~eV and 10.5~eV~\cite{Rescigno2016}:
\begin{align}
\mathrm{e^- + NH_3} &\to \mathrm{NH_2^- + H},\\
\mathrm{e^- + NH_3} &\to \mathrm{NH^- + H_2}.
\end{align}
At sub-eV energies, where electrons drift thermally in a TPC under
moderate reduced fields ($E/N \lesssim 5~\mathrm{Td}$), these channels
are energetically closed and the attachment cross section is negligible.
Therefore, NH$_3$ behaves as a \emph{non-electronegative} additive,
in contrast with species such as O$_2$, H$_2$O, or SF$_6$, which exhibit
strong near-zero-eV resonances.

\subsection{Effects of Contaminants on \nhfp}

A priory \nhfp can react in the gas phase with a
neutral impurity $X$ (such as O$_2$, CO$_2$, or H$_2$O) via proton transfer:
\[
\mathrm{NH_4^+ + X \;\to\; XH^+ + NH_3}.
\]

Whether this reaction can occur at thermal energies is determined by the sign of the Gibbs energy associated to this process, which can be easily estimated from tabulated gas-phase basicities (GB), see Table~\ref{tab:PA_IE} \cite{ZF8572}.\footnote {We adopt the standard IUPAC gas-phase convention and define gas-phase basicity as the negative of the Gibbs energy change associated with the reaction $\mathrm{X + H^+ \;\to\; XH^+}$.}

Accordingly, the Gibbs energy change for proton transfer from \nhfp~to a potential impurity is

\[
\Delta G = \mathrm{GB}(NH_3 ) - \mathrm{GB}(X).
\label{eq:H}
\]

Hence
\begin{itemize}
\item If $\mathrm{GB}(X) > \mathrm{GB}(NH_3)$, then $\Delta G < 0$ and
proton transfer can proceed spontaneously.
\item If $\mathrm{GB}(X) < \mathrm{GB}(NH_3)$, then $\Delta G > 0$ and the
reaction cannot occur thermally.
\end{itemize}

Looking at the values in Table~\ref{tab:PA_IE}, it is evident that all the potential impurities (e.g., H$_2$O, CO$_2$, O$_2$, N$_2$, CO, H$_2$, CH$_4$, CH$_3$CN or CH$_3$OH) present a gas-phase basicity significantly smaller than that of NH$_3$; thus proton transfer will be forbidden in the gas phase. Some of them (e.g.,  H$_2$O,  CH$_3$CN or CH$_3$OH) would interact with NH$_4^+$, but only through weak, reversible clustering.  Species with GB $\mathrm >$ GB(NH$_3$), such as certain amines, can potentially react with NH$_4^+$ in gas phase. However, amines are very efficiently suppressed by hot getters, probably at the level of ppt
concentrations. Electron-transfer neutralization of NH$_4^+$ by impurities is energetically disfavored owing to the very low ionization potential of NH$_4$ and is therefore expected to occur with a negligible probability under the conditions considered here.

Tetraphenyl butadiene (TPB, C$_{28}$H$_{22}$) is a large aromatic hydrocarbon, extensively used as a wavelength 
shifter~\cite{Gehman:2011xm, NEXT:2022cmg}. It is a solid at room temperature, with extremely low vapor 
pressure ($\ll 10^{-12}$~bar).  
Its gas--phase concentration in xenon is therefore
negligible. Even if present in trace amounts, its gas-phase basicity is 
presumably comparable to that of other polycyclic aromatics and does not exceed 
$\mathrm{GB(NH_3)}$. It follows that TPB should not affect the drift of 
\nhfp\ in the gas phase.

\begin{table}[ht]
  \centering
  \caption{Approximate gas--phase basicities of selected species (298~K). Data from ref.~\cite{ZF8572}.}
  
  \label{tab:PA_IE}
  \begin{tabular}{lcc}
    \hline\hline
    Species & $\mathrm{GB}$ [kJ/mol], 298 K  \\
    \hline
    (CH$_3$)$_3$N   & 918.1 \\
    pyridine        & 898.1 \\
    (CH$_3$)$_2$NH   & 896.5 \\
   \textbf{NH$_3$}           & \textbf{854.0} \\
    CH$_3$CN         &748.0 \\
    C$_6$H$_6$            &725.4 \\
    CH$_3$OH        & 724.5  \\
    H$_2$O           & 660.0 \\
    CO              & 562.8  \\
   CH$_4$          & 520.0 \\
   CO$_2$          & 515.8 \\
    N$_2$           & 464.5 \\
    O$_2$           & 396.3  \\
    H$_2$           & 394.7 \\

    \hline\hline
  \end{tabular}
\end{table}

\section{The ion track detector}
\label{sec:itd}

The ion track detector is composed of four main components: the Molecular Ion Detector, the Magnetically-Activated Molecular Apparatus, the Transport and Shelving System, and the Ion Scanning Microscope.

\begin{figure}[!htbp]
    \centering
    \includegraphics[width=0.99\linewidth]{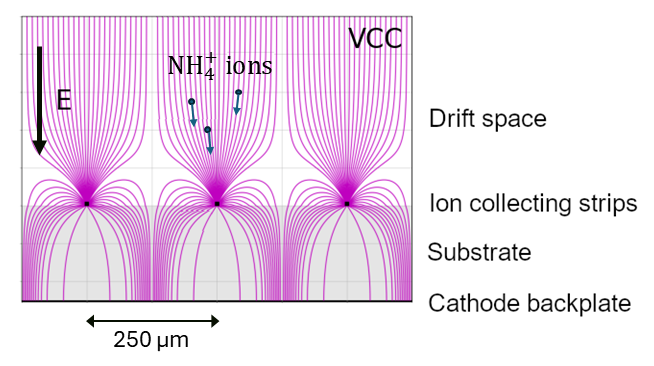}
    \caption{In the MID, ions are focused onto the strips of a VCC microstructure
    similar to that shown in the figure.}
    \label{fig:vcc}
\end{figure}

\vspace{0.5 cm}
\noindent \textbf{The Molecular Ion Detector (MID)} consists of:

\begin{enumerate}
\item A square carrier plate (CP), $10\times10~\mathrm{cm}^2$, large enough to amply contain an ion track  (the typical track length at 30~bar is $\sim 5$~cm, thus about 10 cm size). The CP moves along a rail by magnetic actuation. 
 \item A molecular target (MT), based on a VCC (Virtual Cathode Chamber)~\cite{Capeans1997VCC,Breskin_2022} (figure \ref{fig:vcc}). The VCC consists of a substrate made of resistive material, a conductive film in the back face of the substrate acting as a cathode, and anode strips in the front face. Electrostatic focusing is used to guide the incoming ions onto the active strip regions of the detector, increasing the local ion density at the sensor surface and thereby improving both capture efficiency and signal-to-noise while minimizing the required readout area. The strips are 10 cm long and 50~\si{\micro\meter} wide and are spaced at a pitch of 250~\si{\micro\meter}. The molecular sensors are deposited on top of the anode strips. The MT can be inserted (removed) in a holder fixed to the CP.
\end{enumerate}

\begin{figure}[!htbp]
    \centering
    \includegraphics[scale=0.60]{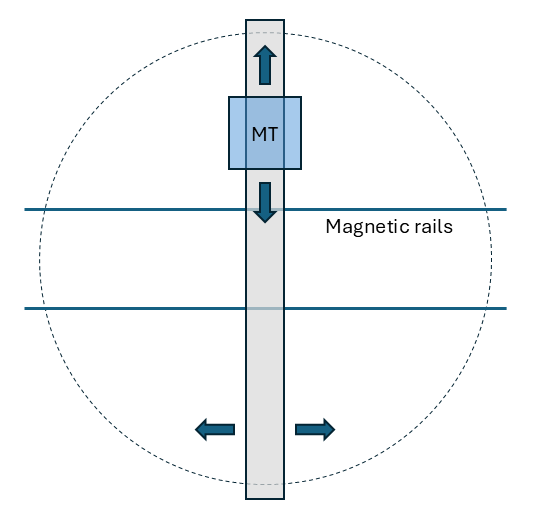}
    \caption{A sketch of the Magnetically Activated Molecular Apparatus (MAMA).}
    \label{fig:mama}
\end{figure}

\noindent \textbf{The Magnetically Activated Molecular Apparatus (MAMA)} is shown in figure \ref{fig:mama}. The simplest realization of the device is a guide that moves left to right on magnetically activated rails, and an MID that moves up to down, also by magnetic activation, on the guide. In practice, the device will contain two or more guides, and each guide will contain two MIDs, to reduce the distance traveled by the MID to intercept the ion. 

\vspace{0.5 cm}

\noindent \textbf{The Transport and Shelving System (TSS)} has the following roles:
\begin{enumerate}
\item It can place and/or remove the MT in the MID. Each time a trigger activates and data is collected, the MT is removed from the MID (after the device returns to its parking position), and a new MT is placed in the MID.
\item It takes the MT to the microscope system for scanning, after a trigger. 
\item It stores the MT in a shelf after scanning. 
\end{enumerate}

\noindent \textbf{The Ion Scanning Microscope (ISM)} consists of a microscope coupled via optical fibers to an excitation laser. The microscope includes a pressure-resistant objective connected to a CCD camera, similar to those developed in the context of the Barium Tagging R\&D Program ~\cite{NEXT:2024tbp} and the necessary instrumentation to perform a fast scanning of the molecular target. 

Consider now a \bbonu\ decay occurring at coordinates $(x,y,z)$ in the ITACA detector. For illustration, take $z = 100$~cm (e.g, in the middle of the TPC).
The electron track is collected by the SiPMs instrumenting the anode approximately 1~ms after $t_0$. Afterwards, one has 2.5 seconds to measure the event energy 
and perform a rough reconstruction of the topology (0.5 seconds if the event occurs at the edge of the fiducial volume, $L=190$~cm). 

An event is classified as a \bbonu\ candidate if:
\begin{enumerate}
\item The event energy lies in the ROI (e.g.\ $\pm 25$~keV around \Qbb).
\item A single track (with no disconnected energy depositions) is reconstructed. 
\end{enumerate}

If both conditions are fulfilled, then:

\begin{enumerate}
    \item The event barycenter is computed in the anode plane, 
    $P = (x_b, y_b, z_a)$.
    \item The barycenter is projected to the detection region, located near the
    cathode at $\mathrm{z = z_c}$, yielding 
    $P' = (x_b, y_b, z_c)$.
    \item The MAMA system positions the MID in $P'$, captures the ion track and then is retracted to its parking position.
    \item The TSS takes the MT from its holder in the MID and introduces a new MT. Then transports the MT to the ISM where is scanned. After scanning the MT is placed in a storage shelf.
  \item The ISM scans the MT and reconstructs the track. 
 \end{enumerate}

\subsection{Molecular sensors}

The same techniques under development for \batp\ tagging could be applied here for 
$\mathrm{NH}_4^{+}$ sensing, e.g. the use of a fluorescent chemical sensor offering a clear luminescent response to the ions. 

Concerning the type of sensors to be employed, fortunately, the development of luminescent sensors for quaternary ammonium cations is a mature area of research. Its development is driven by the ubiquity of these ions in many biologically active molecules and by their participation in several biological processes ($\mathrm{NH}_4^{+}$ is a common cellular metabolite and key for the correct cellular pH balance) \cite{ZF8531}. An overview of the different families of sensors already developed for these ions reveals crown-ethers, aza-crown ethers and related cryptands~\cite{ZF8537} among the most recurrent binding motives~\cite{ZF8531}. Actually, the affinity of these macrocyclic receptors for quaternary ammonium cations is well established since the first reports on crown ethers and the early developments of host-guest chemistry~\cite{ZF8535,ZF8538,ZF8541}. In these examples, the recognition of ammonium ions relies on the presence of hydrogen bonds between the strongly polarized N-H bonds and the lone pairs of the N or O atoms in the receptor, helped by the appropriate steric complementarity (dictated by the size and flexibility of the macrocyclic receptor), being 18-crown-6 type the optimal size and shape for binding  $\mathrm{NH}_4^{+}$ or primary ammonium ions~\cite{ZF8537,ZF8533}.

\begin{figure}[!htbp]
    \centering
    \includegraphics[width=1\linewidth]{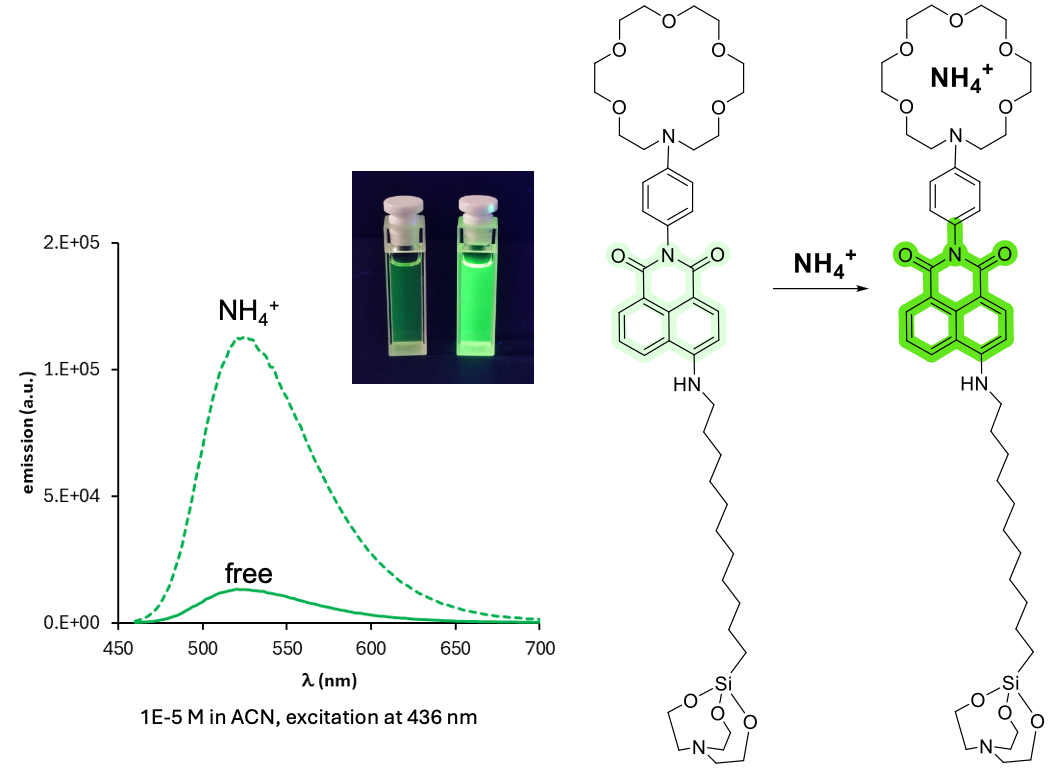}
    \caption{Emission spectra of the free and chelated species (with \nhfp) of NAPH3  (Naphthalimide) sensors. }
    \label{fig:naph}
\end{figure}

Notably, 1-aza-18-crown-6 ethers (and related compounds) are one of the sensors extensively studied within the NEXT Barium Tagging R\&D Program. Therefore, all the know-how already developed can be directly transferred to this new configuration (and vice versa). Having established azacrown ethers as binding units, one can construct off/on, bicolor or time-resolved luminescence sensors depending on the nature of the fluorophore employed. Considering a turning-on response as the most simple sensing mechanism, a naphthalimide derivative (NAPH3, see Figure~\ref{fig:naph}), formerly studied for \batp\  tagging, can be used as a representative example. In this sensor, in the off state (no ion present) the luminescence is quenched due to a Photoinduced Electron Transfer mechanism (PET) from the azacrown ether to the naphthalimide moiety. This process is deactivated upon ion binding.

\begin{figure}[!htbp]
    \centering
    \includegraphics[width=1\linewidth]{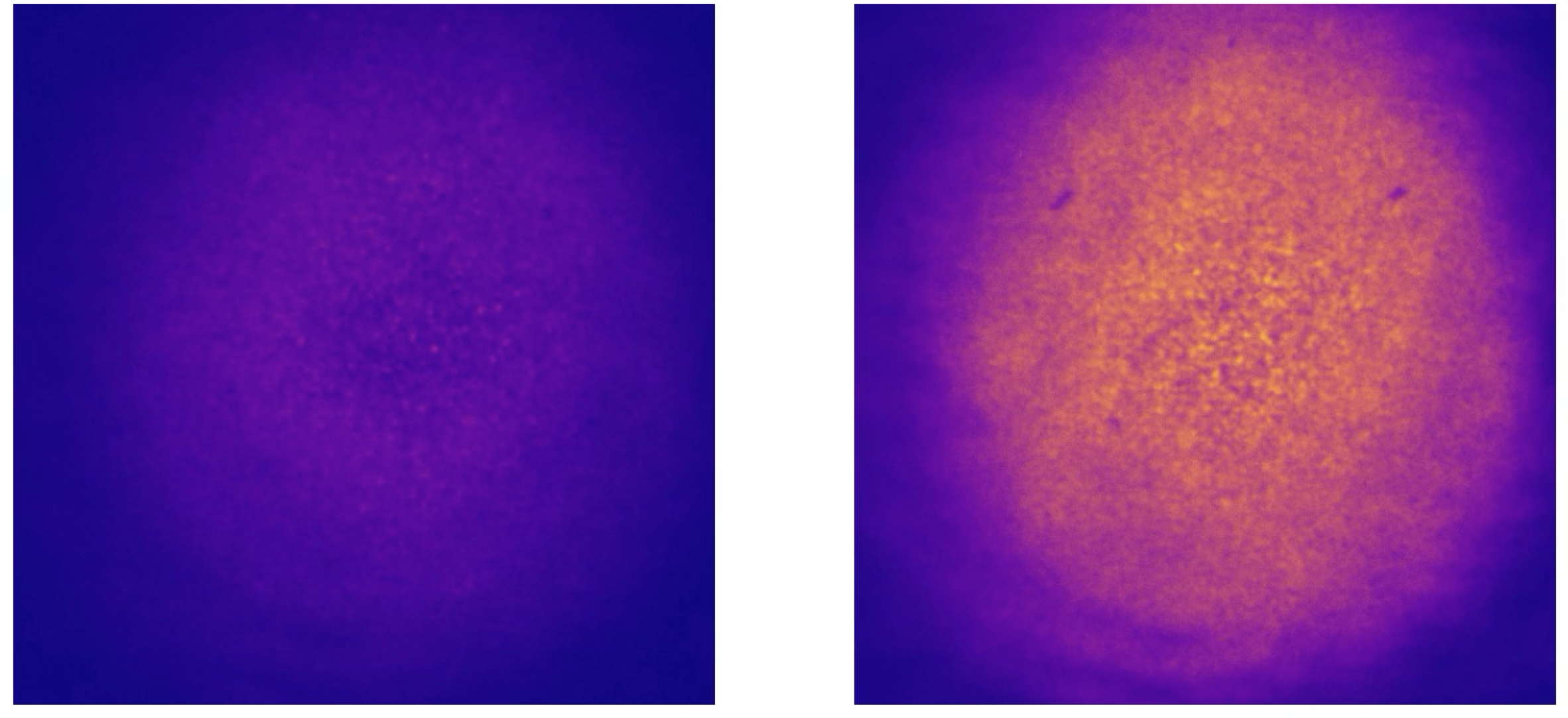}
    \caption{Left panel: response to laser interrogation of a molecular target containing a layer of NAPH3 sensors functionalized in a coverslip. Light emission is very weak, since the PET mechanism quenches fluorescence. Right panel: response to laser interrogation after chelating the molecular target with \nhfp\ ions, showing a strong enhancement of light emission. }
    \label{fig:ms}
\end{figure}

Figure \ref{fig:ms} shows a prototype of a molecular target, containing a layer of NAPH3 sensors functionalized in a coverslip. The response to laser interrogation of the MT is shown before chelating the molecular sensors with \nhfp\ (left panel) and after (right panel), showing a strong enhancement of the light emission and illustrating the soundness of the concept.

\subsection{Measuring the ion track}

The ion track is scanned with a microscope coupled to a suitable excitation laser, a pressure-resistant objective of high NA (0.95) and a high-resolution camera. 

\begin{figure}[!htbp]
    \centering
    \includegraphics[width=1\linewidth]{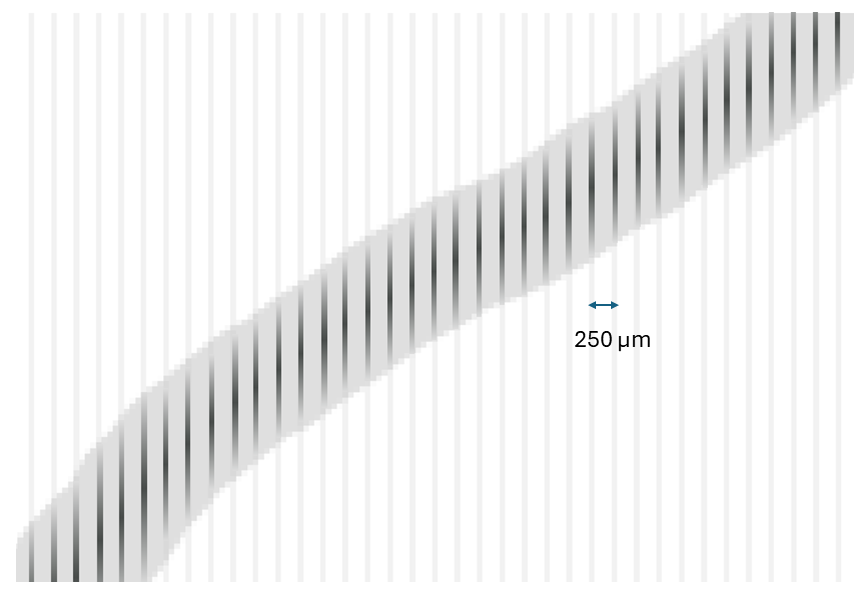}
    \caption{A segment of the track being sampled by the strips in the VCC. }
    \label{fig:strips}
\end{figure}

Figure \ref{fig:strips} illustrates how a track segment is scanned in the VCC. The first step is to project the electron track envelope into the VCC. Then the microscope moves along the strips, recording frames of $50 \times 250$~\si{\micro\meter}$^2$. Consider a typical track in the VCC. The track length is about 50 mm and its width about 4 mm (for a $\sigma_t = 1$~mm). Suppose that the track is parallel to the strips. Then, its envelope contains about 16 strips, and scanning each strip requires 200 frames, resulting in a total of 3200 frames. If the track is perpendicular to the strips, the number of strips is the same. 

The molecular sensor layers (MSL) are deposited on each strip. 
The MSL is characterized by two parameters. Its surface density ($\rho_s$), and the fraction of free molecules that respond to the laser excitation ($f_b$). Both parameters are being investigated in the context of the Barium-tagging R\&D. For a fully packed monolayer $\rho_s \sim 10^6$ molecules/\si{\micro\meter}$^2$, but such a dense arrangement may result in collective effects, and a density of $\rho_s \sim 10^5$ molecules/\si{\micro\meter}$^2$ (molecules spaced about 3 nm) appears to be a much better option. Depending on the structure of MSL (which can be designed co-functionalizing sensors with passive components), densities of $\rho_s \sim 10^4$ molecules/\si{\micro\meter}$^2$ may be feasible, hopefully keeping high ion trapping efficiency.

Reducing as much as possible the fraction of molecules that respond to excitation light when not chelated with the incoming ion, $f_b$, is also an active area of research. PET sensors, on the other hand can yield background fractions below $f_b \sim 10^{-3}$ ~\cite{Niu2023}. 

For illustration purposes, we have run a simulation with the following parameters.

\begin{itemize}
\item Example sensor: NAPH3 (Naphthalimide). 
\item MID as described above: frames of $50 \times 250$ \si{\micro\meter}$^2$.
\item Scanning time per frame: 0.1 s. 
\item Laser power: 10 mW.
\item $\rho_s \times f_b \sim 10^2$ molecules/\si{\micro\meter}$^2$ (for instance,
$\rho_s \sim 10^5$ molecules/\si{\micro\meter}$^2$ and $f_b \sim 10^{-3}$).
\end{itemize}

\begin{figure}[!htbp]
    \centering
    \includegraphics[width=1\linewidth]{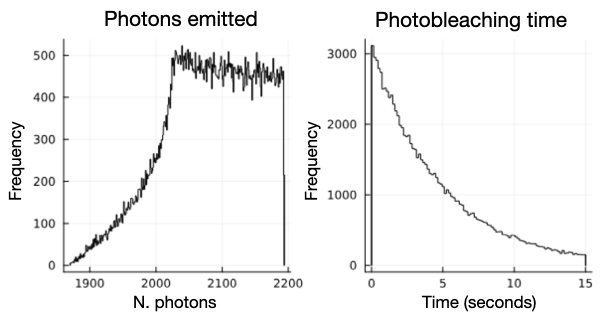}
    \caption{Left: Number of photons per active molecule emitted during the exposition time (0.1 second) in each pin; right: photobleaching time. }
    \label{fig:nphot}
\end{figure}

Figure \ref{fig:nphot} shows the number of photons per active molecule emitted during the exposition time (0.1 second), which is chosen to be small compared with photobleaching time.  The asymmetry in the distribution of the number of emitted photons is related with the non-uniformity of the beam (simulated as a gaussian beam) illumination of the strip. 

\begin{figure*}[!htbp]
    \centering
    \includegraphics[width=0.9\linewidth]{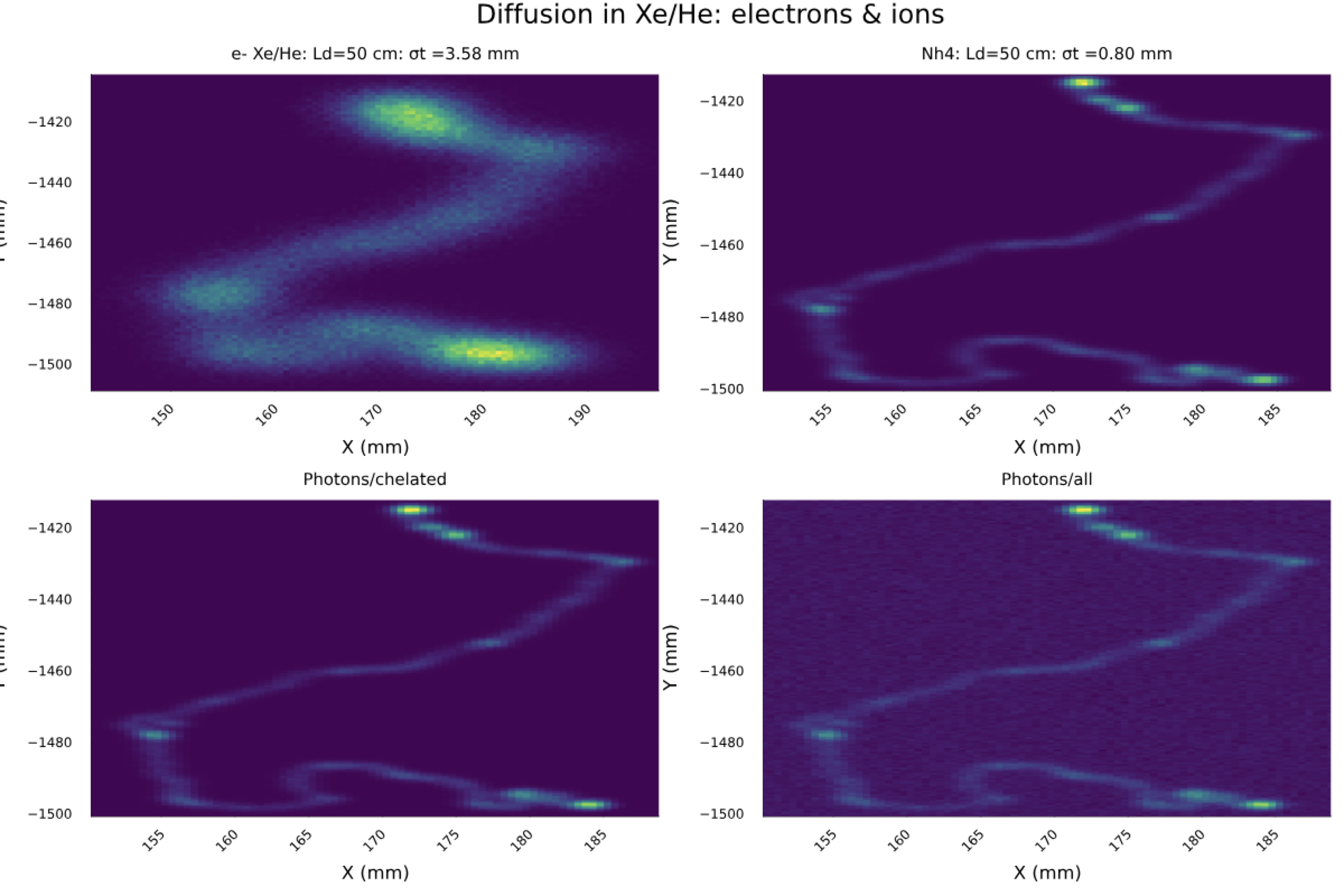}
    \caption{Upper row: electron and ions tracks corresponding to a \bbonu\ event after diffusion of 50 cm; bottom row: reconstruction of the ion track. The bottom left panel shows the image obtained recording the photons emitted only by the chelated molecule.  The bottom right panel shows the image obtained recording the photons emitted by the chelated molecules as well as the fraction of active molecules in the layer.}
    \label{fig:trkrec}
\end{figure*}

Figure \ref{fig:trkrec} shows one example of our simulations. The upper row: shows the electron and ions tracks corresponding to a \bbonu\ event after diffusion of 50 cm. The bottom row shows the reconstruction of the ion track. In the bottom left panel we record only the light produced by the chelated molecules (thus, the fraction of active background molecules is forced to zero). In the bottom right panel we show the image obtained recording the photons emitted by the chelated molecules as well as the fraction of active molecules in the layer. Notice that, although the background noise is visible, the main features of the track, and in particular the clean structure of the blobs, remain unchanged. 

It appears, therefore, possible to measure the ion track with high signal to noise. The time associated to the measurement  if of the order of 300 s, which is perfectly affordable since the scanning of the ion track only occurs for \bbonu\ candidates. 

Given the slow drift velocity of the ions, gas motion associated with continuous purification, thermal gradients, or mechanical actuation could, in principle, affect the ion signal in two ways: by introducing uncertainties in the ion arrival time at the detector and by distorting the reconstructed ion topology. In the ITACA concept, the timing aspect is mitigated by event-driven gating, whereby the ion-collection window is opened only for events that pass the energy selection and the prompt electron-track topology criteria, thereby strongly suppressing pile-up from unrelated ions. Concerning topology, gas recirculation in large high-pressure xenon TPCs is typically operated at low flow rates designed to ensure smooth, quasi-laminar conditions~\cite{NEXT:2025yqw}. Under such conditions, gas motion is expected to induce primarily coherent displacements or mild distortions of the ion track rather than topology scrambling, manifesting as a modest degradation of spatial resolution rather than a loss of topological discrimination.

\section{Enhancement of the sensitivity}
\label{sec:sensi}

The measurement of the ion track will enhance the overall sensitivity of the GXeEL technology along two complementary lines. On one hand, the reduced diffusion permits a better discrimination between near-by floating energy depositions (for example the characteristic 30 keV X-ray emitted in \BI\ decays) and the main track, and on the other, it improves the discrimination between ``single electrons'' and ``double electrons'', since the two blobs are easier to recognize. This enhanced topological separation is particularly relevant for suppressing backgrounds that intrinsically produce single-electron recoils, including those arising from elastic scattering of solar neutrinos, which constitute a background for next-generation ton-scale xenon detectors~\cite{KamLAND-Zen:2024eml}. In addition, topology plays a crucial role in mitigating backgrounds from charged-current solar neutrino interactions on $^{136}$Xe, which produce $^{136}$Cs followed by $\beta$ decays with endpoint energies near the ROI~\cite{Ejiri:2013jda}.

To quantify the improvement in sensitivity relative to that of a GXeEL TPC operating with pure Xe and a Xe/He gas mixture, signal ($0\nu\beta\beta$) and background (individual gamma rays of energy 2.447 MeV) events were simulated uniformly in the active region of the detector. The resulting energy depositions were recorded on a 3D grid of ``voxels'' for input to a Sparse Convolutional neural network \cite{Graham2018}. The network was trained to classify the events into signal and background using approximately 88,500 of each event type in the training set, 50,000 in the validation set, and 20,000 in the test set. This process was repeated for voxel sizes of $1\times1\times1$~mm$^{3}$, $3.5\times3.5\times3.5$~mm$^{3}$, and $10\times10\times10$~mm$^3$ to model the diffusion and reconstruction accuracy in the case of NH$_3$, a GXeEL TPC filled with a Xe/He mixture, and with pure Xe, respectively.

\begin{figure}[!ht]
    \centering
    \includegraphics[width=1\linewidth]{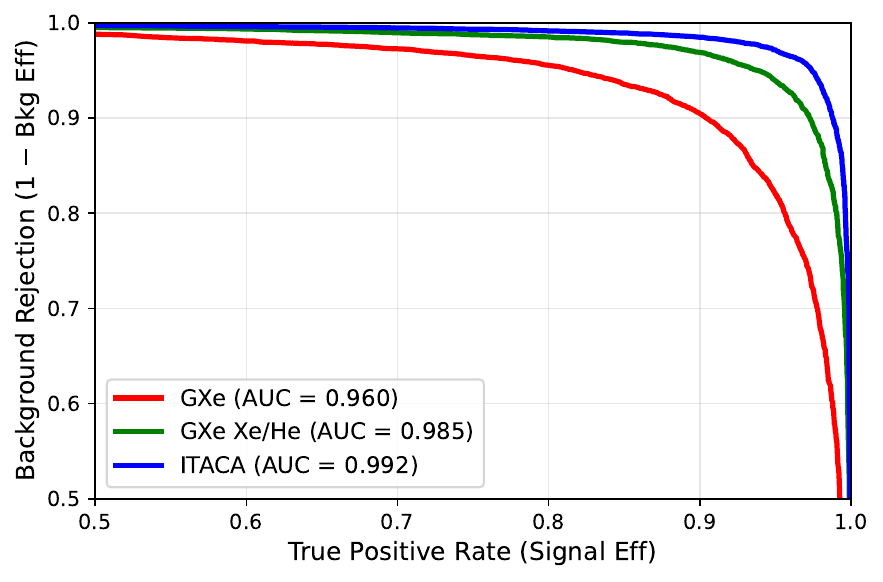}
    \caption{ROC curves for the ITACA detector (blue) and a GXe detector operating with pure Xe (red) and a Xe/He mixture (green). }
    \label{fig:roc}
\end{figure}

Figure \ref{fig:roc} shows the ROC curves obtained in this study.
Using only the electron track yields an area under the curve (AUC) of 0.96 in pure xenon. Including the ion track information improves the AUC to 0.992, corresponding to a factor-of-5 gain in sensitivity, defined as the ratio of \(1-\mathrm{AUC}\) between the two cases. A further improvement arises from the enhanced separation between floating gamma interactions and the main track. In particular, we find an additional factor-of-4 background rejection for \BI\ gamma rays ---the most dangerous background due to their proximity to \Qbb--- leading to an overall \(\times20\) improvement. 

\section{Conclusions}\label{s:conclusions}
\label{sec:conclu}
Measurement of ion tracks with a diffusion of order 1~mm appears feasible in a GXeEL detector. Implementing this concept requires extending the current technology to include controlled trace concentrations of NH$_3$ and an ion-detection system located in the cathode region. 

In addition, the proposed technique is potentially compatible with barium tagging. Since Ba$^{2+}$ ions are expected to drift at a velocity different from that of NH$_4^+$~\cite{Bainglass:2018odn}, a sequential collection scheme is in principle possible in which the NH$_4^+$ ion track is first recorded on the molecular target and, subsequently, the same target is translated by a small distance such that the delayed Ba$^{2+}$ ion is intercepted on an unoccupied region of the sensor surface. Such a sequential scheme could enable the combination of ion-track topology with barium tagging, thereby providing an additional handle for background rejection, even in scenarios where the barium-tagging efficiency is limited.

The additional discrimination provided by the sharply defined ion track further has the potential to reduce the backgrounds by an additional order of magnitude compared with the standard GXeEL technology. This enhanced rejection capability is crucial to achieving the ultra-low background levels required to probe the normal neutrino-mass hierarchy. 

\begin{acknowledgements}
The authors would like to acknowledge support from the following agencies and institutions: the European Research Council (ERC) under Grant Agreement No. 951281-BOLD, the MCIN/AEI of
Spain and ERDF A way of making Europe under grant PID2024-161553NB-C51, the Dpt. of Science, Universities and Innovation of the Basque Government through the 2025 grant to the DIPC Neutrino Physics Laboratory. The experimental part of this work was supported by project  PID2022-139760NB-I00 funded by MCIN/AEI/10.13039/501100011033 and FEDER A way of making Europe.
JR acknowledges support from the Generalitat Valenciana of Spain under grant CIDEXG/2023/16. ME acknowledges support from the Basque Government under grant PRE\_2025\_2\_0146. Discussions with our colleagues in the NEXT collaboration and the BOLD R\&D program are warmly acknowledged.  
\end{acknowledgements}
\bibliographystyle{unsrturl}
\bibliography{pool/NextRefs}

@article{Capeans1997VCC,
  author    = {Cape{\'a}ns, M. and Dominik, W. and Hoch, M. and Ropelewski, L.
               and Sauli, F. and Shekhtman, L. and Sharma, A.},
  title     = {The virtual cathode chamber},
  journal   = {Nuclear Instruments and Methods in Physics Research Section A},
  volume    = {400},
  pages     = {17--23},
  year      = {1997},
  publisher = {Elsevier},
}

@article{Niu2023,
  title = {Photoinduced electron transfer ({PeT}) based fluorescent probes for cellular imaging and disease therapy},
  author = {Niu, Huiyu and Liu, Jie and O'Connor, Helen M. and Gunnlaugsson, Thorfinnur and James, Tony D. and Zhang, Hua},
  journal = {Chemical Society Reviews},
  volume = {52},
  number = {7},
  pages = {2322--2357},
  year = {2023},
  publisher = {Royal Society of Chemistry},
  doi = {10.1039/D1CS01097B},
  url = {https://pubs.rsc.org/en/content/articlehtml/2023/cs/d1cs01097b}
}

@article{nextcollaboration2025demonstrationsubpercentenergyresolution,
      title="{Demonstration of Sub-Percent Energy Resolution in the NEXT-100 Detector}", 
      author={M. Pérez Maneiro and others},
      collaboration = "NEXT",
      year={2025},
      eprint={2511.02467},
      archivePrefix={arXiv},
      primaryClass={physics.ins-det},
}

@article{ABEDI2014101,
title = {Investigation of temperature, electric field and drift-gas composition effects on the mobility of NH4+ ions in He, Ar, N2, and CO2},
journal = {International Journal of Mass Spectrometry},
volume = {370},
pages = {101-106},
year = {2014},
issn = {1387-3806},
doi = {https://doi.org/10.1016/j.ijms.2014.06.014},
url = {https://www.sciencedirect.com/science/article/pii/S1387380614002206},
author = {Azra Abedi and Laleh Sattar and Mahtab Gharibi and Larry A. Viehland}
}

@article{Breskin_2022,
   title={Novel electron and photon recording concepts in noble-liquid detectors},
   volume={17},
   ISSN={1748-0221},
   url={http://dx.doi.org/10.1088/1748-0221/17/08/P08002},
   DOI={10.1088/1748-0221/17/08/p08002},
   number={08},
   journal={Journal of Instrumentation},
   publisher={IOP Publishing},
   author={Breskin, A.},
   year={2022},
   month=aug, pages={P08002} }

@article{Chen1999_NH3_UV_PSS,
  author  = {Chen, F. Z. and Judge, D. L. and Wu, C. Y. Robert and Caldwell, John},
  title   = "{Low and room temperature photoabsorption cross sections of NH$_3$ in the UV region}",
  journal = {Planetary and Space Science},
  year    = {1999},
  volume  = {47},
  pages   = {261--266},
  doi     = {10.1016/S0032-0633(98)00074-9}
}

@article{jones2022ionfluorescencechamberifc,
      title="{The Ion Fluorescence Chamber (IFC): A new concept for directional dark matter and topologically imaging neutrinoless double beta decay searches}", 
      author={B. J. P. Jones and F. W. Foss and J. A. Asaadi and E. D. Church and J. deLeon and E. Gramellini and O. H. Seidel and T. T. Vuong},
      year={2022},
      eprint={2203.10198},
      archivePrefix={arXiv},
      primaryClass={physics.ins-det},
}

@article{Conaway1987,
  author    = {Conaway, Wayne E. and Ebata, Takayuki and Zare, Richard N.},
  title     = "{Vibrationally state‐selected reactions of ammonia ions. III. NH$_3^+$ + NH$_3$ $\rightarrow$ NH$_4^+$ + NH$_2$}",
  journal   = {Journal of Chemical Physics},
  volume    = {87},
  number    = {6},
  pages     = {3453--3460},
  year      = {1987},
  doi       = {10.1063/1.453084}
}

@article{mason1988tables,
  title={Tables of Properties Useful in the Estimation of Ion-Neutral Interaction Energies},
  author={Mason, EA and McDaniel, EW},
  journal={Transport Properties of Ions in Gases, Chapter Appendix III},
  pages={531--540},
  year={1988}
}

@article{Levandier2010,
  author    = {D. J. Levandier and  Y. K. Chiu},
  title     = "{A guided ion beam study of the reactions of Xe$^+$ and Xe$_2^+$ with NH$_3$}",
  journal   = {Journal of Chemical Physics},
  volume    = {133},
  pages     = {154304},
  year      = {2010},
  doi       = {10.1063/1.3496668}
}

@article{Rescigno2016,
  author       = {Rescigno, Thomas N. and Trevisan, Carlos S. and Orel, Ann E. and Slaughter, Daniel S. and Adaniya, Hiroshi and Belkacem, Ali and Weyland, Marvin and Dorn, Alexander and McCurdy, C. W.},
  title        = {Dynamics of dissociative electron attachment to ammonia},
  journal      = {Phys. Rev. A},
  year         = {2016},
  volume       = {93},
  number       = {5},
  pages        = {052704},
  doi          = {10.1103/PhysRevA.93.052704},
  url          = {https://link.aps.org/doi/10.1103/PhysRevA.93.052704}
}

@article{ZF8546,
   author = {Auria-Luna, Fernando and others},
   title = {Supramolecular chemistry in solution and solid–gas interfaces: synthesis and photophysical properties of monocolor and bicolor fluorescent sensors for barium tagging in neutrinoless double beta decay},
   journal = {RSC Appl. Interfaces},
collaboration = {NEXT},
   volume = {2},
   number = {1},
   pages = {185-199},
   DOI = {10.1039/D4LF00227J},
   url = {http://dx.doi.org/10.1039/D4LF00227J},
   year = {2025},
   type = {Journal Article}
}

@article{ZF8538,
   author = {Cram, Donald J. and Cram, Jane M.},
   title = {Host-Guest Chemistry},
   journal = {Science},
   volume = {183},
   number = {4127},
   pages = {803-809},
   DOI = {10.1126/science.183.4127.803},
   url = {https://doi.org/10.1126/science.183.4127.803},
   year = {1974},
   type = {Journal Article}
}

@article{ZF8535,
   author = {Newcomb, Martin and Timko, Joseph M. and Walba, David M. and Cram, Donald J.},
   title = {Host-guest complexation. 3. Organization of pyridyl binding sites},
   journal = {J. Am. Chem. Soc.},
   volume = {99},
   number = {19},
   pages = {6392-6398},
   ISSN = {0002-7863},
   DOI = {10.1021/ja00461a035},
   url = {https://doi.org/10.1021/ja00461a035},
   year = {1977},
   type = {Journal Article}
}

@article{ZF8541,
   author = {Pedersen, Charles J.},
   title = {Cyclic polyethers and their complexes with metal salts},
   journal = {J. Am. Chem. Soc.},
   volume = {89},
   number = {26},
   pages = {7017-7036},
   ISSN = {0002-7863},
   DOI = {10.1021/ja01002a035},
   url = {https://doi.org/10.1021/ja01002a035},
   year = {1967},
   type = {Journal Article}
}

@article{ZF8533,
   author = {Seilkop, Austin G. and Odoh, Amaechi S. and Coradi, Nicholas J. and Wright, Jacob I. and Barroso, Jorge and Kim, Byoungmoo},
   title = {Ammonium-Binding Bifunctional Aza-Crown Ether Catalysts for Substrate-Selective Hydroxyl Functionalization},
   journal = {J. Org. Chem.},
   volume = {89},
   number = {18},
   pages = {13338-13344},
   ISSN = {0022-3263},
   DOI = {10.1021/acs.joc.4c01498},
   url = {https://doi.org/10.1021/acs.joc.4c01498},
   year = {2024},
   type = {Journal Article}
}

@article{ZF8537,
   author = {Shen, Min and Pan, Tingting and Chen, Yonghao and Ning, Juewei and Su, Fengyu and Tian, Yanqing},
   title = {Highly selective fluorescent sensor for ammonium ions},
   journal = {Sens. Diagn.},
   volume = {3},
   number = {1},
   pages = {79-86},
   DOI = {10.1039/D3SD00128H},
   url = {http://dx.doi.org/10.1039/D3SD00128H},
   year = {2024},
   type = {Journal Article}
}

@article{ZF8531,
   author = {Späth, A. and König, B.},
   title = {Molecular recognition of organic ammonium ions in solution using synthetic receptors},
   journal = {Beilstein J. Org. Chem.},
   pages = {Apr 6;6:32},
   year = {2010},
   type = {Journal Article}
}

@book{Mason1988,
  author       = {Mason, E. A. and McDaniel, E. W.},
  title        = {Transport Properties of Ions in Gases},
  publisher    = {John Wiley \& Sons},
  address      = {New York},
  year         = {1988},
  isbn         = {978-0-471-84812-8}
}

@article{doi:10.1021/acssensors.4c01892,
author = {Aranburu, Ane I. and Elorza, Mikel and Valle, Pablo R.G. and Pazos, Ariadna and Brodolin, Alexey and Herrero-Gómez, Pablo and Barcelon, J. Eduardo and Molina-Terriza, Gabriel and Monrabal, Francesc and Rogero, Celia and Cossío, Fernando P. and Gómez-Cadenas, Juan Jos{\'e} and Tonnel{\'e}, Claire and Freixa, Zoraida},
collaboration = "NEXT",
title = "{Iridium-Based Time-Resolved Luminescent Sensor for Ba2+ Detection}",
journal = {ACS Sensors},
volume = {10},
number = {4},
pages = {2487-2498},
year = {2025},
doi = {10.1021/acssensors.4c01892},
    note ={PMID: 40213995},

URL = { 
    
        https://doi.org/10.1021/acssensors.4c01892
    
    

},
eprint = { 
    
        https://doi.org/10.1021/acssensors.4c01892
    
    

}

}

@article{NEXT:2025yqw,
    author = "Adams, C. and others",
    collaboration = "NEXT",
    title = "{The NEXT-100 Detector}",
    eprint = "2505.17848",
    archivePrefix = "arXiv",
    primaryClass = "physics.ins-det",
    reportNumber = "FERMILAB-PUB-25-0365-V",
    month = "5",
    year = "2025"
}

@article{McDonald_2019,
    doi = {10.1088/1748-0221/14/08/P08009},
    url = {https://doi.org/10.1088/1748-0221/14/08/P08009},
    year = {2019},
    volume = {14},
    number = {08},
    pages = {P08009},
    eprint = {1902.05544},
    author = {McDonald, A.D. and others},
    title = {Electron drift and longitudinal diffusion in high pressure xenon-helium gas mixtures},
    journal = {JINST}
}

@article{Soleti:2023sac,
    author = "Soleti, S. R.",
    collaboration = "NEXT",
    title = "{Towards a fiber barrel detector for next-generation high-pressure gaseous xenon TPCs}",
    eprint = "2312.05567",
    archivePrefix = "arXiv",
    primaryClass = "physics.ins-det",
    doi = "10.1088/1748-0221/19/04/C04042",
    journal = "JINST",
    volume = "19",
    number = "04",
    pages = "C04042",
    year = "2024"
}

@article{Graham2018,
    author={Graham, Benjamin and van der Maaten, Laurens},
    title={Submanifold Sparse Convolutional Networks},
    year={2017},
    eprint = {1706.01307},
    archivePrefix = {arXiv},
    primaryClass = {cs.CV}
}

@article{Martoff:2000wi,
    author = "Martoff, C. J. and Snowden-Ifft, D. P. and Ohnuki, T. and Spooner, N. and Lehner, M.",
    title = "{Suppressing drift chamber diffusion without magnetic field}",
    doi = "10.1016/S0168-9002(99)00955-9",
    journal = "Nucl. Instrum. Meth. A",
    volume = "440",
    pages = "355--359",
    year = "2000"
}

@article{Snowden-Ifft:1999reu,
    author = "Snowden-Ifft, D. P. and Martoff, C. J. and Burwell, J. M.",
    title = "{Low pressure negative ion drift chamber for dark matter search}",
    eprint = "astro-ph/9904064",
    archivePrefix = "arXiv",
    doi = "10.1103/PhysRevD.61.101301",
    journal = "Phys. Rev. D",
    volume = "61",
    pages = "101301",
    year = "2000"
}

@article{LEGEND2025,
    title="{First Results on the Search for Lepton Number Violating Neutrinoless Double Beta Decay with the LEGEND-200 Experiment}", 
    author={H. Acharya and others},
    collaboration = "{LEGEND}",
    year={2025},
    eprint={2505.10440},
    archivePrefix={arXiv},
}

@article{Ejiri:2013jda,
    author = "Ejiri, H. and Elliott, S. R.",
    title = "{Charged current neutrino cross section for solar neutrinos, and background to $\beta\beta(0\nu)$ experiments}",
    eprint = "1309.7957",
    archivePrefix = "arXiv",
    primaryClass = "nucl-ex",
    reportNumber = "LA-UR-13-27310",
    doi = "10.1103/PhysRevC.89.055501",
    journal = "Phys. Rev. C",
    volume = "89",
    number = "5",
    pages = "055501",
    year = "2014"
}

@article{NEXT:2022cmg,
    author = "Haefner, J. and others",
    collaboration = "NEXT",
    title = "{Reflectance and fluorescence characteristics of PTFE coated with TPB at visible, UV, and VUV as a function of thickness}",
    eprint = "2211.05024",
    archivePrefix = "arXiv",
    primaryClass = "physics.ins-det",
    reportNumber = "FERMILAB-PUB-22-845-ND-SCD",
    doi = "10.1088/1748-0221/18/03/P03016",
    journal = "JINST",
    volume = "18",
    number = "03",
    pages = "P03016",
    year = "2023"
}

@article{KamLANDZen2024,
    title = "{Search for Majorana Neutrinos with the Complete KamLAND-Zen Dataset}", 
    author = "Abe, S. and others",
    year = {2024},
    eprint = "2406.11438",
    doi = {10.48550/arXiv.2406.11438},
    archivePrefix = "arXiv"
}

@article{Avignone2008,
    author = "Avignone III, F. T. and Elliott, S. R. and Engel, J.",
    title = "{Double beta decay, Majorana neutrinos, and neutrino mass}",
    eprint = "0708.1033",
    archivePrefix = "arXiv",
    doi = "10.1103/RevModPhys.80.481",
    journal = "Rev. Mod. Phys.",
    volume = "80",
    pages = "481",
    year = "2008"
}

@article{Dolinski2019,
    author = "Dolinski, Michelle J. and Poon, Alan W.P. and Rodejohann, Werner",
    title = "{Neutrinoless Double-Beta Decay: Status and Prospects}", 
    journal= "Ann. Rev. Nucl. Part. Sci.",
    year = "2019",
    volume = "69",
    number = "Volume 69, 2019",
    pages = "219-251",
    doi = "10.1146/annurev-nucl-101918-023407",
    url = "https://www.annualreviews.org/content/journals/10.1146/annurev-nucl-101918-023407",
    publisher = "Annual Reviews",
    issn = "1545-4134",
    eprint = "1902.04097"
}

@article{NEXT:2024tbp,
    author = "Byrnes, N. K. and others",
    collaboration = "NEXT",
    title = "{Fluorescence imaging of individual ions and molecules in pressurized noble gases for barium tagging in $^{136}$Xe}",
    eprint = "2406.15422",
    archivePrefix = "arXiv",
    primaryClass = "physics.ins-det",
    reportNumber = "FERMILAB-PUB-24-0802-PPD",
    doi = "10.1038/s41467-024-54872-0",
    journal = "Nature Commun.",
    volume = "15",
    number = "1",
    pages = "10595",
    year = "2024"
}

@article{Akiyama:2025dia,
    author = "Akiyama, Shinichi and others",
    title = "{In-situ high voltage generation with Cockcroft-Walton multiplier for xenon gas time projection chamber}",
    eprint = "2501.08554",
    archivePrefix = "arXiv",
    primaryClass = "physics.ins-det",
    month = "1",
    year = "2025"
}

@article{NEXT:2020amj,
    author = "Adams, C. and others",
    collaboration = "NEXT",
    title = "{Sensitivity of a tonne-scale NEXT detector for neutrinoless double beta decay searches}",
    eprint = "2005.06467",
    archivePrefix = "arXiv",
    primaryClass = "physics.ins-det",
    doi = "10.1007/JHEP08(2021)164",
    journal = "JHEP",
    volume = "2021",
    number = "08",
    pages = "164",
    year = "2021"
}

@article{NEXT:2022ita,
    author = "Herrero-G\'omez, P. and others",
    collaboration = "NEXT",
    title = "{Ba$^{+2}$ ion trapping using organic submonolayer for ultra-low background neutrinoless double beta detector}",
    eprint = "2201.09099",
    archivePrefix = "arXiv",
    primaryClass = "hep-ex",
    reportNumber = "FERMILAB-PUB-22-614-ND-SCD",
    doi = "10.1038/s41467-022-35153-0",
    journal = "Nature Commun.",
    volume = "13",
    number = "1",
    pages = "7741",
    year = "2022"
}

@article{KamLAND-Zen:2024eml,
    author = "Abe, S. and others",
    collaboration = "KamLAND-Zen",
    title = "{Search for Majorana Neutrinos with the Complete KamLAND-Zen Dataset}",
    eprint = "2406.11438",
    archivePrefix = "arXiv",
    primaryClass = "hep-ex",
    month = "6",
    year = "2024"
}

@article{Gomez-Cadenas:2023vca,
    author = "G\'omez-Cadenas, Juan Jos\'e and Mart\'\i{}n-Albo, Justo and Men\'endez, Javier and Mezzetto, Mauro and Monrabal, Francesc and Sorel, Michel",
    title = "{The search for neutrinoless double-beta decay}",
    doi = "10.1007/s40766-023-00049-2",
    journal = "Riv. Nuovo Cim.",
    volume = "46",
    number = "10",
    pages = "619--692",
    year = "2023"
}

@article{rivilla_fluorescent_2020,
	title = {Fluorescent bicolour sensor for low-background neutrinoless double $\beta$ decay experiments},
	volume = {583},
	copyright = {2020 The Author(s), under exclusive licence to Springer Nature Limited},
	issn = {1476-4687},
	url = {https://www.nature.com/articles/s41586-020-2431-5},
	doi = {10.1038/s41586-020-2431-5},
	language = {en},
	number = {7814},
	urldate = {2021-03-16},
	journal = {Nature},
	author = {Rivilla, Iv\'an and Aparicio, Borja and Bueno, Juan M. and Casanova, David and Tonnel\'e, Claire and Freixa, Zoraida and Herrero, Pablo and Rogero, Celia and Miranda, Jos\'e I. and Mart\'inez-Ojeda, Rosa M. and Monrabal, Francesc and Olave, Be\~nat and Schafer, Thomas and Artal, Pablo and Nygren, David and Coss\'io, Fernando P. and G\'omez-Cadenas, Juan J.},
	month = jul,
	year = {2020},
	note = {Number: 7814
Publisher: Nature Publishing Group},
	pages = {48--54}
}

@Article{Alvarez:2012hh,
  author        = {\'Alvarez, V. and others},
  title         = {{Near-Intrinsic Energy Resolution for 30 to 662 keV Gamma Rays in a High Pressure Xenon Electroluminescent TPC}},
  journal       = {Nucl.\ Instrum.\ Meth.},
  year          = {2012},
  volume        = {A708},
  pages         = {101-114},
  archiveprefix = {arXiv},
  collaboration = {NEXT},
  doi           = {10.1016/j.nima.2012.12.123},
  eprint        = {1211.4474},
  primaryclass  = {physics.ins-det},
  slaccitation  = {%%CITATION = ARXIV:1211.4474;%%},
}

@article{Henriques:2018tam,
      author         = "Henriques, C. A. O. and others",
      title          = "{Electroluminescence TPCs at the Thermal Diffusion
                        Limit}",
      collaboration  = "NEXT",
      journal        = "JHEP",
      volume         = "01",
      year           = "2019",
      pages          = "027",
      doi            = "10.1007/JHEP01(2019)027",
      eprint         = "1806.05891",
      archivePrefix  = "arXiv",
      primaryClass   = "physics.ins-det",
      reportNumber   = "FERMILAB-PUB-18-290-CD",
      SLACcitation   = "%%CITATION = ARXIV:1806.05891;%%"
}

@article{Simon:2017pck,
      author         = "Sim\'on, A. and others",
      title          = "{Application and performance of an ML-EM algorithm in
                        NEXT}",
      collaboration  = "NEXT",
      journal        = "JINST",
      volume         = "12",
      year           = "2017",
      number         = "08",
      pages          = "P08009",
      doi            = "10.1088/1748-0221/12/08/P08009",
      eprint         = "1705.10270",
      archivePrefix  = "arXiv",
      primaryClass   = "physics.ins-det",
      reportNumber   = "FERMILAB-PUB-17-219-CD",
      SLACcitation   = "%%CITATION = ARXIV:1705.10270;%%"
}

@article{Henriques:2017rlj,
      author         = "Henriques, C. A. O. and others",
      title          = "{Secondary scintillation yield of xenon with sub-percent
                        levels of CO$_2$ additive for rare-event detection}",
      collaboration  = "NEXT",
      journal        = "Phys. Lett.",
      volume         = "B773",
      year           = "2017",
      pages          = "663-671",
      doi            = "10.1016/j.physletb.2017.09.017",
      eprint         = "1704.01623",
      archivePrefix  = "arXiv",
      primaryClass   = "physics.ins-det",
      reportNumber   = "FERMILAB-PUB-17-225-CD-ND",
      SLACcitation   = "%%CITATION = ARXIV:1704.01623;%%"
}

@article{Felkai:2017oeq,
      author         = "Felkai, R. and others",
      title          = "{Helium–Xenon mixtures to improve the topological
                        signature in high pressure gas xenon TPCs}",
      journal        = "Nucl. Instrum. Meth.",
      volume         = "A905",
      year           = "2018",
      pages          = "82-90",
      doi            = "10.1016/j.nima.2018.07.013",
      eprint         = "1710.05600",
      archivePrefix  = "arXiv",
      primaryClass   = "physics.ins-det",
      reportNumber   = "FERMILAB-PUB-17-688-CD-ND",
      SLACcitation   = "%%CITATION = ARXIV:1710.05600;%%"
}

@article{Bainglass:2018odn,
      author         = "Bainglass, E. and Jones, B. J. P. and Foss, F. W. and
                        Huda, M. N. and Nygren, D. R.",
      title          = "{Mobility and Clustering of Barium Ions and Dications in
                        High Pressure Xenon Gas}",
      journal        = "Phys. Rev.",
      volume         = "A97",
      year           = "2018",
      number         = "6",
      pages          = "062509",
      doi            = "10.1103/PhysRevA.97.062509",
      eprint         = "1804.01169",
      archivePrefix  = "arXiv",
      primaryClass   = "physics.ins-det",
      SLACcitation   = "%%CITATION = ARXIV:1804.01169;%%"
}

@article{Alvarez:2011my,
      author         = "\'Alvarez, V. and others",
      title          = "{The NEXT-100 experiment for neutrinoless double beta
                        decay searches (Conceptual Design Report)}",
      collaboration  = "NEXT",
      year           = "2011",
      eprint         = "1106.3630",
      archivePrefix  = "arXiv",
      primaryClass   = "physics.ins-det",
      SLACcitation   = "%%CITATION = ARXIV:1106.3630;%%",
}

@article{Alvarez:2012haa,
      author         = "\'Alvarez, V. and others",
      title          = "{NEXT-100 Technical Design Report (TDR): Executive
                        Summary}",
      collaboration  = "NEXT",
      journal        = "JINST",
      volume         = "7",
      pages          = "T06001",
      doi            = "10.1088/1748-0221/7/06/T06001",
      year           = "2012",
      eprint         = "1202.0721",
      archivePrefix  = "arXiv",
      primaryClass   = "physics.ins-det",
      SLACcitation   = "%%CITATION = ARXIV:1202.0721;%%",
}

@article{Alvarez:2012as,
      author         = "\'Alvarez, V. and others",
      collaboration = "NEXT",
      title          = "{Radiopurity control in the NEXT-100 double beta decay
                        experiment: procedures and initial measurements}",
      year           = "2012",
      journal        = "JINST",
      volume         = "8",
      pages          = "T01002",
      doi            = "10.1088/1748-0221/8/01/T01002",
      eprint         = "1211.3961",
      archivePrefix  = "arXiv",
      primaryClass   = "physics.ins-det",
      SLACcitation   = "%%CITATION = ARXIV:1211.3961;%%",
}

@article{Alvarez:2012hu,
      author         = "\'Alvarez, V. and others",
      title          = "{Ionization and scintillation response of high-pressure
                        xenon gas to alpha particles}",
      collaboration  = "NEXT",
      year           = "2013",
      eprint         = "1211.4508",
      archivePrefix  = "arXiv",
      primaryClass   = "physics.ins-det",
      journal        = "JINST",
      volume         = "1305",
      pages          = "P05025",
      doi            = "10.1088/1748-0221/8/05/P05025",
      SLACcitation   = "%%CITATION = ARXIV:1211.4508;%%",
}

@article{Alvarez:2012xda,
      author         = "\'Alvarez, V. and others",
      title          = "{Initial results of NEXT-DEMO, a large-scale prototype of the NEXT-100 experiment}",
      collaboration  = "NEXT",
      journal        = "JINST",
      volume         = "8",
      pages          = "P04002",
      doi            = "10.1088/1748-0221/8/04/P04002",
      year           = "2013",
      eprint         = "1211.4838",
      archivePrefix  = "arXiv",
      primaryClass   = "physics.ins-det",
      SLACcitation   = "%%CITATION = ARXIV:1211.4838;%%",
}

@article{Ferrario:2015kta,
      author         = "Ferrario, P. and others",
      title          = "{First proof of topological signature in the high
                        pressure xenon gas TPC with electroluminescence
                        amplification for the NEXT experiment}",
      collaboration  = "NEXT",
      journal        = "JHEP",
      volume         = "01",
      year           = "2016",
      pages          = "104",
      doi            = "10.1007/JHEP01(2016)104",
      eprint         = "1507.05902",
      archivePrefix  = "arXiv",
      primaryClass   = "physics.ins-det",
      reportNumber   = "FERMILAB-PUB-15-648-CD-ND",
      SLACcitation   = "%%CITATION = ARXIV:1507.05902;%%"
}

@article{Gehman:2011xm,
      author         = "Gehman, V. M. and Seibert, S. R. and Rielage, K. and
                        Hime, A. and Sun, Y. and Mei, D. M. and Maassen, J. and
                        Moore, D.",
      title          = "{Fluorescence Efficiency and Visible Re-emission Spectrum
                        of Tetraphenyl Butadiene Films at Extreme Ultraviolet
                        Wavelengths}",
      journal        = "Nucl. Instrum. Meth.",
      volume         = "A654",
      year           = "2011",
      pages          = "116-121",
      doi            = "10.1016/j.nima.2011.06.088",
      eprint         = "1104.3259",
      archivePrefix  = "arXiv",
      primaryClass   = "astro-ph.IM",
      reportNumber   = "LA-UR-11-10447",
      SLACcitation   = "%%CITATION = ARXIV:1104.3259;%%"
}

@article{Simon:2018vep,
      author         = "Sim\'on, A. and others",
      title          = "{Electron drift properties in high pressure gaseous
                        xenon}",
      collaboration  = "NEXT",
      year           = "2018",
      journal        = "JINST",
      volume         = "13",
      pages          = "P07013",
      eprint         = "1804.01680",
      archivePrefix  = "arXiv",
      primaryClass   = "physics.ins-det",
      SLACcitation   = "%%CITATION = ARXIV:1804.01680;%%"
}

@article{McDonald:2017izm,
      author         = "McDonald, A. D. and others",
      title          = "{Demonstration of Single Barium Ion Sensitivity for
                        Neutrinoless Double Beta Decay using Single Molecule
                        Fluorescence Imaging}",
      collaboration  = "NEXT",
      journal        = "Phys. Rev. Lett.",
      volume         = "120",
      year           = "2018",
      number         = "13",
      pages          = "132504",
      doi            = "10.1103/PhysRevLett.120.132504",
      eprint         = "1711.04782",
      archivePrefix  = "arXiv",
      primaryClass   = "physics.ins-det",
      reportNumber   = "FERMILAB-PUB-17-521-ND",
      SLACcitation   = "%%CITATION = ARXIV:1711.04782;%%"
}

@article{Jones:2016qiq,
    author = "Jones, B. J. P. and McDonald, A. D. and Nygren, D. R.",
    title = "{Single Molecule Fluorescence Imaging as a Technique for Barium Tagging in Neutrinoless Double Beta Decay}",
    eprint = "1609.04019",
    archivePrefix = "arXiv",
    primaryClass = "physics.ins-det",
    doi = "10.1088/1748-0221/11/12/P12011",
    journal = "JINST",
    volume = "11",
    number = "12",
    pages = "P12011",
    year = "2016"
}

@article{Thapa:2019zjk,
      author         = "Thapa, P. and Arnquist, I. and Byrnes, N. and Denisenko,
                        A. A. and Foss, F. W. and Jones, B. J. P. and Mcdonald, A.
                        D. and Nygren, D. R. and Woodruff, K.",
      title          = "{Barium Chemosensors with Dry-Phase Fluorescence for
                        Neutrinoless Double Beta Decay}",
      year           = "2019",
      eprint         = "1904.05901",
      archivePrefix  = "arXiv",
      primaryClass   = "physics.ins-det",
      SLACcitation   = "%%CITATION = ARXIV:1904.05901;%%"
}

@article{NEXT:2019oxh,
    author = "Fernandes, A. F. M. and others",
    collaboration = "NEXT",
    title = "{Low-diffusion Xe-He gas mixtures for rare-event detection: Electroluminescence Yield}",
    eprint = "1906.03984",
    archivePrefix = "arXiv",
    primaryClass = "physics.ins-det",
    doi = "10.1007/JHEP04(2020)034",
    journal = "JHEP",
    volume = "04",
    pages = "034",
    year = "2020"
}

@article{ZF7462,
   author = {Freixa, Zoraida and Rivilla, Iván and Monrabal, Francesc and Gómez-Cadenas, Juan J. and Cossío, Fernando P.},
   title = {Bicolour fluorescent molecular sensors for cations: design and experimental validation},
   journal = {Phys. Chem. Chem. Phys.},
   volume = {23},
   number = {29},
   pages = {15440-15457},
   ISSN = {1463-9076},
   DOI = {10.1039/D1CP01203G},
   url = {http://dx.doi.org/10.1039/D1CP01203G},
   year = {2021},
   type = {Journal Article}
}

@article{ZF7997,
   author = {Thapa, Pawan and Byrnes, Nicholas K. and Denisenko, Alena A. and Mao, James X. and McDonald, Austin D. and Newhouse, Charleston A. and Vuong, Thanh T. and Woodruff, Katherine and Nam, Kwangho and Nygren, David R. and Jones, Benjamin J. P. and Foss, Frank W., Jr.},
   title = {Demonstration of Selective Single-Barium Ion Detection with Dry Diazacrown Ether Naphthalimide Turn-on Chemosensors},
   journal = {ACS Sens.},
   volume = {6},
   number = {1},
   pages = {192-202},
   DOI = {10.1021/acssensors.0c02104},
   url = {https://doi.org/10.1021/acssensors.0c02104},
   year = {2021},
   type = {Journal Article}
}

@article{ZF8572,
   author = {Hunter, Edward P. L. and Lias, Sharon G.},
   title = {Evaluated Gas Phase Basicities and Proton Affinities of Molecules: An Update},
   journal = {Journal of Physical and Chemical Reference Data},
   volume = {27},
   number = {3},
   pages = {413-656},
   ISSN = {0047-2689},
   DOI = {10.1063/1.556018},
   url = {https://doi.org/10.1063/1.556018},
   year = {1998},
   type = {Journal Article}
}

\end{document}